\documentclass[%
a4paper,prl,final,superscriptaddress,twocolumn, floatfix
]{revtex4-2}

\usepackage{graphicx}
\usepackage{dcolumn}
\usepackage{bm}
\usepackage{siunitx}
\usepackage{tabularx}
\usepackage{latexsym}
\usepackage{changes}
\usepackage{esvect}
\usepackage[colorlinks = true,
            linkcolor = blue,
            urlcolor  = blue,
            citecolor = blue,
            anchorcolor = blue]{hyperref}
\usepackage{mathtools}
\usepackage{xcolor}
\usepackage{soul}
\usepackage{amssymb}
\usepackage{bm}
\usepackage{color}
\usepackage{braket}
\usepackage{dsfont}
\usepackage{wasysym}
\usepackage{xfrac}
\usepackage{braket}
\usepackage{textgreek}
\usepackage{ulem}
\usepackage{cleveref}

\usepackage{makecell}

\begin{document}
\setcounter{figure}{0}

\title{Implementation of scalable suspended superinductors}
\author{Christian J{\"u}nger}
\thanks{These two authors contributed equally.}

\affiliation{Department of Physics, University of California, Berkeley, CA 94720, USA}
\affiliation{Applied Mathematics and Computational Research Division, Lawrence Berkeley National Laboratory, Berkeley, California 94720, USA}

\author{Trevor Chistolini}
\thanks{These two authors contributed equally.}
\affiliation{Department of Physics, University of California, Berkeley, CA 94720, USA}

\author{Long B. Nguyen}
\thanks{Current address: AWS Center for Quantum Computing, Pasadena, CA 91125, USA}
\affiliation{Department of Physics, University of California, Berkeley, CA 94720, USA}
\affiliation{Applied Mathematics and Computational Research Division, Lawrence Berkeley National Laboratory, Berkeley, California 94720, USA}

\author{Hyunseong Kim}
\affiliation{Department of Physics, University of California, Berkeley, CA 94720, USA}

\author{Larry Chen}
\affiliation{Department of Physics, University of California, Berkeley, CA 94720, USA}

\author{\\ Thomas Ersevim}
\affiliation{Department of Physics, University of California, Berkeley, CA 94720, USA}

\author{William Livingston}
\affiliation{Department of Physics, University of California, Berkeley, CA 94720, USA}

\author{Gerwin Koolstra}
\affiliation{Department of Physics, University of California, Berkeley, CA 94720, USA}
\affiliation{Applied Mathematics and Computational Research Division, Lawrence Berkeley National Laboratory, Berkeley, California 94720, USA}

\author{David I. Santiago}
\affiliation{Department of Physics, University of California, Berkeley, CA 94720, USA}
\affiliation{Applied Mathematics and Computational Research Division, Lawrence Berkeley National Laboratory, Berkeley, California 94720, USA}

\author{Irfan Siddiqi}
\affiliation{Department of Physics, University of California, Berkeley, CA 94720, USA}
\affiliation{Applied Mathematics and Computational Research Division, Lawrence Berkeley National Laboratory, Berkeley, California 94720, USA}
\affiliation{Materials Science Division, Lawrence Berkeley National Laboratory, Berkeley, California 94720, USA}

\date{\today}

\begin{abstract}

Superinductors have become a crucial component in the superconducting circuit toolbox, playing a key role in the development of more robust qubits. Enhancing the performance of these devices can be achieved by suspending the superinductors from the substrate, thereby reducing stray capacitance. Here, we present a fabrication framework for constructing superconducting circuits with suspended superinductors in planar architectures. To validate the effectiveness of this process, we systematically characterize both resonators and qubits with suspended arrays of Josephson junctions, ultimately confirming the high quality of the superinductive elements. In addition, this process is broadly compatible with other types of superinductors and circuit designs. Our results not only pave the way for scalable novel superconducting architectures but also provide the primitive for future investigation of loss mechanisms associated with the device substrate.

\end{abstract}

\maketitle

\noindent\emph{Introduction} --
\label{sec:intro}
Multiple superconducting circuit designs demand inductive elements with impedance larger than the resistance quantum $R_Q$ = 6.5 k$\Omega$, where such elements are referred to as superinductors \cite{Manucharyan_fluxonium}. These include novel qubits such as fluxonium~\cite{Manucharyan_fluxonium,pop2014coherent,Nguyen_high_coherence_fluxonium,zhang2021universal}, 0 - $\pi$ \cite{brooks2013protected,Gyenis2021}, bi-fluxon \cite{Kalashnikov_bifluxon}, $\cos 2\phi$ \cite{bell2014protected,larsen2020protected,smith2022magnifying,dodge2023hardware}, and blochnium~\cite{Pechenezhskiy_blochnium}. Realizing this element is typically challenging, noting the intrinsically low vacuum impedance $Z_\mathrm{vac}=\sqrt{\mu_0/\epsilon_0}=377~\Omega$, and various methods have been employed.
Presently, superinductors are conventionally constructed from materials with high kinetic inductance, such as amorphous Al~\cite{Maleeva_CQED_grAl, Gruenhaupt2019}, ultra-thin Al~\cite{Kalacheva_Qi_SC_res_treatment}, NbN~\cite{franca2023nbn} or TiN~\cite{Koolstra_e_He_2024} films, nanowires \cite{Hazard_nanowire_fluxonium}, or Josephson metamaterials \cite{Castellanos-Beltran_metamaterial_amplifier,Ranadive_Kerr_reversal_amplifier}. Etched geometric spiral superconducting wires may also yield high-impedance elements~\cite{peruzzo2020surpassing}.

With such devices in mind, as new designs and parameter spaces are continually explored, systems may require elements of increasingly higher inductance or greater impedance to be practically realized. However, when utilizing materials whose inductance scales with length, it is not always viable to continually increase its inductance or impedance by merely increasing its length. For one, the element's capacitance may scale similarly, thereby limiting the achievable device impedance. Furthermore, scaling its length would decrease the frequency of associated parasitic modes that could interfere with device modes. 
To tackle these problems and expand achievable bounds, one possible route is to reduce an element's capacitance by suspending it from the substrate, thereby removing the most significant dielectric. Such a strategy has been implemented using silicon membranes to achieve geometric superinductors \cite{peruzzo2020surpassing}. In relevance to superconducting qubits, Ref.~\cite{Pechenezhskiy_blochnium} drastically reduced the spurious capacitance of a Josephson junction (JJ) array by suspending it above the substrate, achieving the ``hyperinductance regime" with impedance $R > 200$ k$\Omega$.

\begin{figure*}[t!]
\centering
\includegraphics[width=17cm]{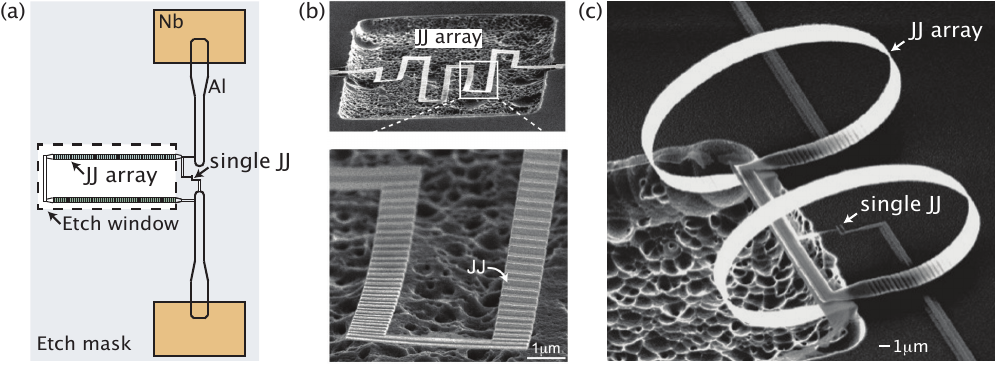}
\caption{Device fabrication. (a) Etching method schematic for selectively suspended fluxonium. The etch mask is shown in grey, with the open etch window for the JJ array. (b) Scanning-electron microscope (SEM) image of a suspended resonator consisting of 500 JJs. The Nb capacitor paddles remain on the substrate (not seen). (c) SEM image of suspended fluxonium with selectively suspended JJ array (250 JJs). The single JJ, which lies outside of the etch window, remains on the Si substrate.}
	\label{fig:fab}
\end{figure*}

In order to incorporate suspended elements into large-scale quantum processors, the fabrication method must not introduce significant loss or damage other components. The current state of the art etching process inevitably causes the release or suspension of all Al structures \cite{Pechenezhskiy_blochnium}. While the technique was used to reduce the capacitance associated with a JJ array, etching the substrate in general could introduce further loss to components that were not intended to be suspended \cite{Dunsworth_bandaid_2017}. This may be partially responsible for the relatively lower device lifetime compared to on-substrate fluxonium counterparts~\cite{Nguyen_high_coherence_fluxonium, Somoroff_ms_fluxonium}.
In addition, such a method limits the selectivity of the process because the entire chip is exposed to the etchant. For instance, the utilized etchant XeF$_2$ is incompatible with Nb, Ta, TiN, and NbN, which are metals often used in high quality resonators and ground plane components \cite{Murray_mat_rev_2021}. 

To address these concerns, we present a versatile technique for selective substrate etching that enables the suspension of arbitrary components in aluminum-based planar superconducting circuits, facilitating scalable novel architectures with suspended superconducting qubits. To demonstrate the efficacy of our protocol and assess the effects of etching, we fabricate and characterize several resonators and fluxonium qubits. Notably, our method systematically reduces stray capacitance and increases inductance without compromising the quality of the measured devices. Furthermore, the energy relaxation times of the suspended resonators and qubits produced with this process are on par with the state-of-the-art. As loss mechanisms in superconducting circuits such as dielectric loss~\cite{chen2024phonon,zhang2024acceptor} and quasiparticle poisoning~\cite{Catelani_decoherence_qp, Glazman_qp_lec_notes, wilen2021correlated,iaia2022phonon,diamond2022distinguishing, connolly2024coexistence} are often linked to defects in or originating from the substrate, we envision future investigation and potential mitigation of these loss mechanisms in suspended devices constructed using this framework. 

\noindent\emph{Device Fabrication} --
\label{sec:fab}
To fabricate these suspended devices, we follow the general fabrication procedure of our wafer scale multi-qubit devices \cite{Hashim2024}, ensuring that the process is similarly scalable. Hence, the ground plane consists of Nb and is fabricated on a 6-inch intrinsic Si wafer, using in-house developed cleaning methods \cite{Altoe2022}. 
The superinductor is constructed from an array of JJs, which is fabricated by a double angle evaporation of Al-AlOx-Al in the Dolan style \cite{Dolan1977}. 

Subsequently, an etch mask (see dashed line in Fig.~\ref{fig:fab}(a)) is lithographically defined across all devices on the wafer. This enables the selective etching of the silicon substrate and prompts the individual lifting of the JJ array, by releasing the strain that is present at the Al - Si interface and results in the array curling. Here, we use $\mathrm{XeF}_2$ as a reactive etchant, an ingredient used extensively in the MEMS community \cite{Winters_XeF2, Chang_XeF2, Chan_XeF2} and recently employed in superconducting circuits \cite{Pechenezhskiy_blochnium}. However, the crucial distinction in our study is that we can selectively etch individual areas of the devices, where the vast majority of the Si surface remains un-etched and pristine. 
As a further consequence of this selectivity, this method allows the integration of alternative materials such as Nb for the resonator ground plane to yield components of high quality factor, which would otherwise be rapidly etched by $\mathrm{XeF}_2$ if not protected by the etch mask. 

This etch mask is subsequently removed by oxygen ashing, as the suspended devices are too fragile to withstand the surface tension of usual solvents used for resist removal. This integrates suspended structures into the wafer-scale fabrication of various devices such as the resonators and qubits shown in Fig. \ref{fig:fab}(b) and (c). In the following, we present a comprehensive study and compare systematic differences between on-substrate and suspended devices.

\noindent\emph{Suspended Resonators} --
\label{sec:resonators}
We primarily characterize two chips of resonators to evaluate the effects of etching and achievable device quality. Each chip contains five arrays of $100 - 500$ JJs in series, in $100$ junction increments, shunted by capacitor paddles and coupled to a central transmission line in an hanger geometry \cite{McRae_res}. The paddles mediate measurement by coupling to the transmission line, and they also modify the device frequencies due to the associated shunting capacitance. While one chip contains suspended resonators, the other contains their on-substrate counterparts. A Nb coplanar waveguide (CPW) resonator is included on each chip as a reference device.

A JJ array can be described as a multi-mode non-linear resonator with self-Kerr ($K_{ii}$) and cross-Kerr ($K_{ij}$) coefficients, with a corresponding Hamiltonian \cite{Masluk_JJ_array, Weissl_Kerr_coefficients}
\begin{equation}\label{eqn:Kerr_Hamiltonian}
    \mathcal{H} = \left( \tilde{\omega}_i - \frac{1}{2}K_{ii}\bar{n}_i - \frac{1}{2}K_{ij}\bar{n}_j \right) a_i^\dagger a_i,
\end{equation}
where $\bar{n}_j$ is the average photon number in mode $j$, $\tilde{\omega}_i=\omega_i - K_{ii}/2 - \sum_j K_{ij}/4$ is a renormalized frequency due to the Kerr coefficients,
\begin{equation}\label{eqn:isolated_array_freq}
    \omega_k= \omega_0 \sqrt{\frac{1-\cos{(\pi k/N)}}{1-\cos{(\pi k/N)} + C_0/(2C_J)}}
\end{equation}
is the bare frequency of an array with $N$ junctions, and $\omega_0 = 1/\sqrt{L_J C_J}$ is the plasma frequency of an individual junction. While we include the cross-Kerr terms in Eq.~\ref{eqn:Kerr_Hamiltonian}, we do not present analysis of them for our devices. The $L_J$ of each array can be extracted by room temperature probing, yielding $0.91$ ($1.10$) nH/junction for on-substrate (suspended) arrays. Notably, such an increase in inductance after etching has also been observed in previous works~\cite{Chu_micromachining}, with possible explanations described later.
Further details on the probing, which helped guide device design, are presented in the Supplementary Information.

\begin{figure}[t!]
\includegraphics[width=\columnwidth]{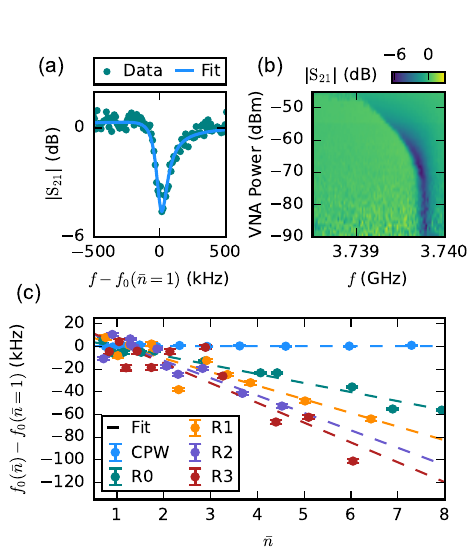}
\caption{Characterization of Josephson junction arrays. (a)~Amplitude response at $\bar{n} \sim 1$, with fit overlaid. (b) Power sweep, demonstrating decreasing frequency with increasing drive powers due to the self-Kerr coefficient. (c) Frequency shift versus average photon number, estimated using our line attenuation calibration. Overlaid linear fits correspond to the self-Kerr coefficients of the arrays.}
	\label{fig:res_characterization}
\end{figure}

As resonators, the arrays can be characterized by their scattering parameters, $S_{21}$, which are measured using a vector network analyzer (VNA) and subsequently used to extract parameters including the resonator frequency and internal quality factor, $Q_i$. Following Eq.~\ref{eqn:Kerr_Hamiltonian}, at low drive powers, the array can be treated as a harmonic oscillator, so it can be fit as shown in Fig.~\ref{fig:res_characterization}(a) using typical methods as in Ref.~\cite{Altoe2022}. 
At approximately single-photon levels, the mean and standard deviation of the internal quality factors of the on-substrate and suspended arrays were $Q_i = (7.4 \pm 1.3) \times 10^4$ and $Q_i = (3.6 \pm 0.9) \times 10^4$, respectively, which are comparable with leading published results on JJ arrays \cite{Masluk_JJ_array}.

Internal and external quality factor values for the individual resonators are shown in the Supplementary Information.
The reduced $Q_i$ of the suspended arrays may largely be attributed to the more gentle cleaning demanded by their fragility. Furthermore, the process of releasing the junction array from the substrate may introduce further loss sources at this new underside metal-air interface, relative to a typically clean metal-substrate interface, and in general alter the device's participation ratios \cite{Wenner_res_sim, Woods_res}.

Due to the Kerr nonlinearity of the JJ arrays, their frequencies should decrease with increasing drive power, as modeled in Eq.~\ref{eqn:Kerr_Hamiltonian}. Furthermore, as shown in Fig.~\ref{fig:res_characterization}(b), the resonator response becomes asymmetric and its linewidth broadens at higher powers due to photon shot noise \cite{Joshi_TiN_Kerr}. For clarity, we include individual traces of resonator responses at varying powers in the Supplementary Information.
We appeal to the self-Kerr coefficient to evaluate whether the suspended arrays still exhibit the expected nonlinear characteristics of JJ arrays. 
Thus, we extract the self-Kerr coefficient of the suspended array's $\omega_{10}$ mode by fitting its frequency versus average photon number $\bar{n}$, as shown in Fig.~\ref{fig:res_characterization}(c). 

To do so, we first identify the input power value corresponding to $\bar{n} \sim 1$, and then linearly fit the resonator frequencies from the previous one and the next ten steps of 1-dBm each.
We calibrate the line attenuation using room temperature measurements, adjusting for the differences in cable attenuation when cold. Further details on our attenuation calibration are included in the Supplementary Information. The resulting self-Kerr coefficients are exhibited in Table~\ref{table:Kerr_coef}. The uncertainties in the line attenuation and the limited signal strength at low $\bar{n}$ prevent a further quantitative analysis of the corresponding Kerr coefficients. However, the trend of increasing self-Kerr coefficients with decreasing junction number follows the expected behavior~\cite{Krupko_Kerr_metamaterial_2018}.

\begingroup
\setlength{\tabcolsep}{6pt}
\renewcommand{\arraystretch}{1.2}
\begin{table}
\centering
\begin{tabular}{|c|c|c|}
 \hline
 Device & Junction Num. & Self-Kerr Coef. ($\text{kHz}/\bar{n}$) \\
 \hline 
 CPW & N/A & $0.0$ \\
 \hline 
 R0 & 400 & $9.0$ \\
 \hline 
 R1 & 300 & $12.0$ \\
 \hline 
 R2 & 200 & $15.4$ \\
 \hline 
 R3 & 100 & $17.5$ \\
 \hline 
\end{tabular}
\caption{Table of the extracted device self-Kerr coefficients of the fundamental mode of the suspended resonators. We utilize our best estimates for line attenuation but acknowledge limitations in the extracted values due to uncertainty. R$j$ where $j=\{0,1,2,3\}$ indicates a JJ array, indexed in ascending frequency order. CPW indicates the coplanar waveguide resonator on the chip.}
\label{table:Kerr_coef}
\end{table}
\endgroup

Having confirmed that the suspended arrays are high quality and exhibit their characteristic nonlinearity, we proceed to investigate if the etching reduces the capacitance to ground, $C_0$. We extract $C_0$ from the array's frequency, employing the Quantum Phase Model (QPM) for a JJ array, including terms from the capacitor paddles~\cite{Fazio_SC_networks, Vool_rev}, described by the Hamiltonian 
\begin{equation}\label{eqn:QPM_Hamiltonian}
    \mathcal{H} = \frac{1}{2}\sum_{i,j} q_i C^{-1}_{i,j} q_j - \sum_{<i,j>} E_{J,<i,j>}\cos(\phi_i - \phi_j),
\end{equation}
which yields Eq. \ref{eqn:isolated_array_freq} in the limit where both endpoints of the array are grounded \cite{Weissl_Kerr_coefficients}. Here, the operators $q_j$ and $\phi_j$ are the number of charge carriers and phase of the island $j$, which obey the commutation relation $[\phi_i, q_j]=2ei\delta_{ij}$, $C_{i,j}$ is the capacitance matrix, and $E_{J,<i,j>}$ is the Josephson energy of the junction between neighboring islands $\{i,j=i+1\}$.
Adapting the method from Ref.~\cite{Weissl_Kerr_coefficients}, we find a mean reduction of approximately $74\%$ in $C_0$ due to the etching process, demonstrating one main goal of the technique. Details on the routine to extract $C_0$ are provided in Supplementary Information. \\

\noindent\emph{Suspended Fluxoniums} --
\label{sec:qubit}
\begin{figure}[h!]
\includegraphics[width=\columnwidth]{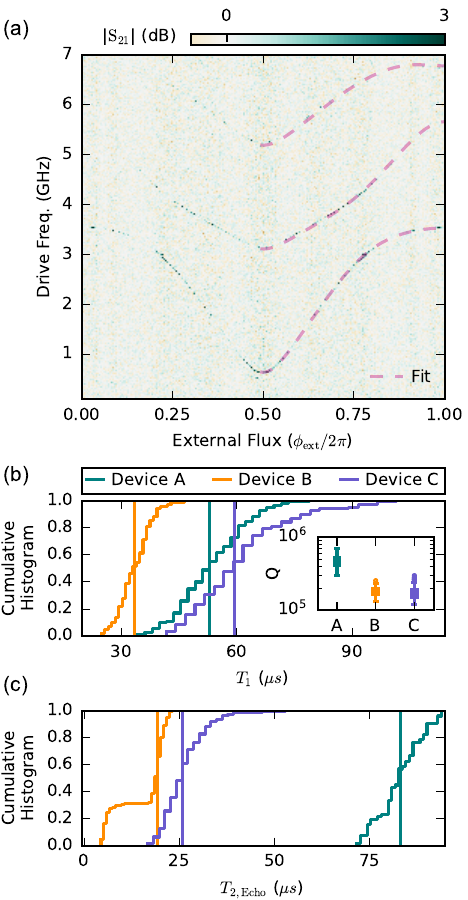}
\caption{Characterization of suspended fluxonium. (a) Transition frequency spectrum of a suspended fluxonium, with extracted $0-1$, $0-2$, and $0-3$ transitions overlaid in magenta over half the spectrum for clarity. (b, c) Cumulative histogram of $T_\mathrm{1}$ ($T_\mathrm{2,Echo}$) measurements for devices. Device A is the conventionally cleaned fluxonium, Device B is the fluxonium on the chip where other devices were etched, and Device C is the suspended fluxonium. Inset in (b) is a boxplot of device quality factors, where whiskers denote $1.5$ times the interquartile range (IQR) from the first and third quartiles. The skewness of Device B's $T_\mathrm{2,Echo}$ data likely arose from a time-dependent TLS during data collection.}
	\label{fig:suspended_qb_char}
\end{figure}
In addition to examining individual arrays, we incorporate a suspended JJ array into a fluxonium type qubit, where a single JJ and capacitor are in parallel on-substrate \cite{Manucharyan_fluxonium, Nguyen_high_coherence_fluxonium, Ding2023}. 
The circuit is described by the Hamiltonian 
\begin{equation}\label{eqn:H_fluxonium}
    \mathcal{H} = 4 E_C n^2 - E_J \cos{\phi} + \frac{1}{2}E_L (\phi+\phi_\mathrm{ext})^2,
\end{equation}
with relevant energy scales $E_C=e^2/(2C)$, $E_J=\Phi_0 I_c/(2\pi)$, and $E_L = (\hbar/2e)^2/L$, corresponding to the capacitance $C$, the small junction critical current $I_c$, and the array inductance $L$. An off-chip coil tunes the external flux $\phi_\mathrm{ext}$. 

We characterize three qubits on two different chips: Device A is a fully on-substrate fluxonium qubit that is cleaned with conventional methods involving solvents, as described in Ref.~\cite{Kreikebaum2020}. Devices B and C are on the same chip, where device B (C) is an on-substrate (partially suspended) fluxonium qubit. The second chip that includes a suspended inductor, in device C, was cleaned without any liquids. Room temperature probing data of the qubits are provided in the Supplementary Information.

Device A's spectrum corresponds to circuit parameters $[E_J, E_C, E_L] = [1.32, 0.93, 0.73]$~GHz. 
At the $\phi_\mathrm{ext}/2\pi = 0.5$ flux bias, $T_\mathrm{1} = 53 \pm 9$ $\mathrm{\mu s}$ corresponds to a quality factor of $Q = 2\pi f T_\mathrm{1} = 4.6\times 10^5$, notably greater than that of the individual arrays, as has been observed in recent work~\cite{Mencia_integer_fluxonium}. The average measured coherence time is $T_\mathrm{2,Echo} = 82 \pm 6$ $\mathrm{\mu s}$. For device B, the extracted parameters are $[E_J, E_C, E_L] = [2.56, 0.96, 0.78]$ GHz, and the measured coherence time scales are $T_\mathrm{1} = 33 \pm 4$ $\mathrm{\mu s}$ ($Q = 1.8\times 10^5$), and $T_\mathrm{2,Echo} = 19 \pm 7$ $\mathrm{\mu s}$. Its associated energy spectrum is included in the Supplementary Information. 
The spectrum of the partially suspended device C is shown in Fig.~\ref{fig:suspended_qb_char}(a), which corresponds to $[E_J, E_C, E_L] = [2.59, 1.01, 0.42]$ GHz. Its average coherence times are $T_\mathrm{1} = 59 \pm 13$ $\mathrm{\mu s}$ ($Q = 1.7\times 10^5$), and $T_\mathrm{2,Echo} = 26 \pm 6$ $\mathrm{\mu s}$. The $T_\mathrm{1}, T_\mathrm{2,Echo}$, and $Q$ data of all three devices are presented in Fig.~\ref{fig:suspended_qb_char}(b,c).

Examining the effects of etching, we find that device C’s characteristic inductance $L$ is 87\% greater than that of device B on the same chip and 75\% greater than that of device A, while devices A and B have similar inductance. This observation suggests that the increase in inductance arises from the etching process, consistent with findings in previous work~\cite{Chu_micromachining}. Notably, a similar effect can be achieved through annealing~\cite{Hertzberg_laser_annealing, Kim_laser_annealing}. We thus believe that the etching process effectively thickens the tunnel barrier, potentially alters the tunnel junction chemistry by introducing fluoride~\cite{Sharma_Al2O3_fluorides}, and might modify the surface through the formation of $\text{AlF}_x$ ~\cite{Roodenko_XeF2_Al_effects}. 

The comparable quality factors of devices B and C indicate that the etching method does not inherently introduce additional impurities. Furthermore, this suggests that the array-substrate interface does not serve as a significant loss channel in fluxonium, such as one where phonons within the substrate might directly couple to the array and constitute a two-level-system bath~\cite{chen2024phonon,zhang2024acceptor} or transmit quasiparticles~\cite{Vepsalainen_qp_ionizing_rad}. On the other hand, the relatively low coherence times in the suspended device can be attributed to dephasing imposed by the fluctuation of thermal photon numbers in the readout resonator, a common problem in the circuit quantum electrodynamics architecture (further details in the Supplementary Information).\\
\noindent\emph{Conclusion} -- We systematically fabricate and characterize superconducting devices with partially suspended superinductors. The extracted self-Kerr coefficients of the suspended resonators exhibit excellent agreement with theory, and they show a dramatic reduction of parasitic capacitance to ground. The suspended fluxonium device reveals a clean spectrum from which we find an $87\%$ increase in inductance compared to its neighboring on-substrate device. Our results imply that the reduction in their quality factors after etching compared to the conventional on-substrate counterparts can be attributed to the gentler cleaning process. Looking forward, more aggressive cleaning could be made compatible with etched devices by using critical point drying (CPD) to remove solvents~\cite{Jafri_CPD_1999, Chistolini_SiN_membrane}. Additionally, the deviation in room-temperature probing resistances of all the fabricated devices are only approximately 10\% (see Supplementary Information). We therefore conclude that the presented method provides a valid route towards the implementation of novel types of qubits that require large inductance elements while minimizing stray capacitance. 

Our selective etching technique integrates seamlessly with current fabrication workflows, offering compatibility with various metals and creating new opportunities for constructing high-impedance superconducting architectures while modifying the quantum system’s noise environment. This approach allows for reliable production of planar hyperinductors, paving the way for scalable, novel qubits such as blochnium~\cite{Pechenezhskiy_blochnium} and the 0-$\pi$ qubit~\cite{brooks2013protected}. Empirical experiments using our protocol could establish optimal processes to suspend other types of superinductors. Moreover, as the investigation of noise mechanisms in superinductors is an active research area, we anticipate systematic exploration of noise sources relevant to devices suspended from the substrate, including dielectric loss~\cite{chen2024phonon,zhang2024acceptor}, $1/f$ flux noise~\cite{Koch_flux_noise, Kumar_flux_noise, braumuller2020characterizing,rower2023evolution}, and quasiparticle poisoning effects~\cite{Catelani_decoherence_qp, Glazman_qp_lec_notes, wilen2021correlated,iaia2022phonon,diamond2022distinguishing, connolly2024coexistence}.

\emph{Correspondence} -- Correspondence should be addressed to C.J. (email: christian.h.juenger@gmail.com) or T.C. (email: trevor\_chistolini@berkeley.edu).\\

\emph{Competing interests} -- The authors declare that they have no competing financial interests.\\

\emph{Author contributions} -- C.J. and T.C. fabricated the devices, measured the samples, and analyzed the results. L.B.N. set up the experimental apparatus and assisted with data acquisition and analysis. L.C. assisted with fabrication. H.K., T.E., and W.L. assisted with operating the dilution refrigerators. G.K. designed and installed the cryogenic wiring. D.I.S. and I.S. supervised the project. C.J., T.C., and L.B.N. wrote the manuscript with input from all authors.\\

\emph{Acknowledgements} --
T.C. acknowledges support from the National Science Foundation Graduate Research Fellowship Program (NSF GRFP) under Grant No. DGE 1752814 and DGE 2146752.
This work is supported by a collaboration between the U.S. DOE and other Agencies. This material is based upon work supported by the U.S. Department of Energy, Office of Science, National Quantum Information Science Research Centers, Quantum Systems Accelerator. This material is based upon work supported in part by the U.S. Army Research Office under grant W911NF-22-1-0258.

\cleardoublepage
\onecolumngrid
\newcommand{\beginsupplement}{
    \setcounter{figure}{0} 
  \setcounter{table}{0}
    \renewcommand{\thetable}{S\arabic{table}}  
  \renewcommand{\thefigure}{S\arabic{figure}}
  \renewcommand{\theHfigure}{Supplement.\thefigure}
}

\section{Supplementary Information \\for ``Implementation of Scalable Suspended Superinductors''}
\beginsupplement

\subsection{Junction Array Wafer Scale Room Temperature Probing Statistics}
\label{supp:RT_probe}

\begin{figure}[h!]
\includegraphics[scale=0.9]{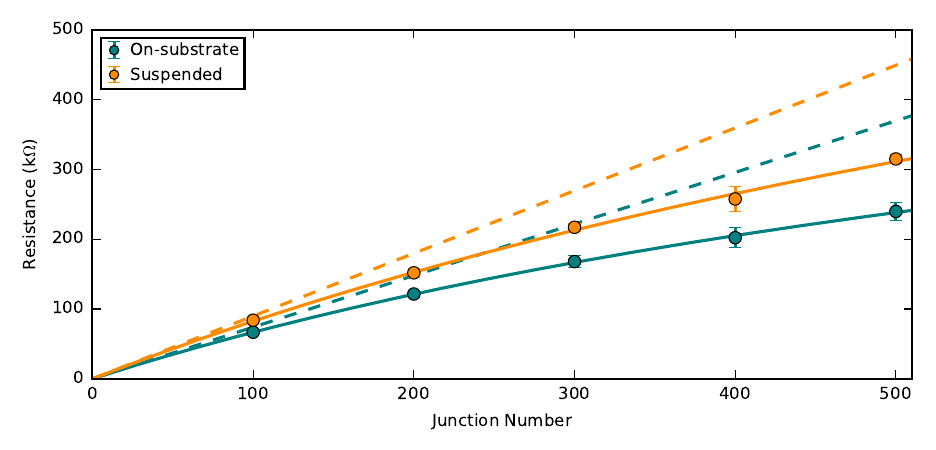}
\caption{Room temperature probe results of the arrays, displaying both on-substrate (teal) and suspended (orange) device results. Error bars for each data point represent the standard deviation of the collected data, with some bars obscured by the point itself. Fits to a parallel resistance model are overlaid in solid lines, accounting for the nonlinear trend with the number of junctions due to the conduction pathway through substrate. Dashed lines indicate the extracted junction resistance.}
	\label{fig:Probe_data}
\end{figure}

Here, we present the results from room temperature, two-point probing of the wafer of junction array dies that the devices in the main text were selected from. Due to the large array resistances, the measured values do not scale linearly with junction number. Instead, they taper off due to the conduction pathway through the substrate becoming non-negligible at high device resistances. To account for this, we fit the probe values to a parallel resistance model of the array and substrate paths, enabling us to identify the isolated junction resistance. Devices are fabricated on a 6-inch Si wafer, as described in the main text. 32 chips are probed, each containing five different lengths of JJ arrays, which constitute the ``on-substrate" data points. The ``suspended" data points represent measurements of three chips drawn from the full wafer that were etched. Each data point is the mean value of those devices, with error bars corresponding to the standard deviation.

In order to translate the room-temperature resistance values to critical currents, we use the Ambegaokar-Baratoff formula \cite{Ambegaokar_tunneling},

\begin{equation}\label{eqn:AB_formula}
    I_c R_n = \frac{\pi \Delta}{2e},
\end{equation}
where $I_c$ is the junction critical current, $R_n$ is the measured normal-state resistance, and $\Delta$ is the superconducting gap. The $I_c$ directly yields the $L_J$ of the junction array. From the fitted lines in Fig. \ref{fig:Probe_data}, we extract values of $L_J = 0.91$ nH for on-substrate junctions and $L_J = 1.10$ nH for suspended junctions. 
The parallel resistance fit also yields substrate resistances of $R = 670$ k$\Omega$ for on-substrate devices and $R=1000$ k$\Omega$ for etched devices. A difference in substrate conductivity between the two device types is expected because the etching occurs between the two pads used for probing contacts.

\subsection{Device Internal and External Quality Factor Statistics}
\label{supp:device_Qi}

\begin{figure}[h!]
\includegraphics[scale=1]{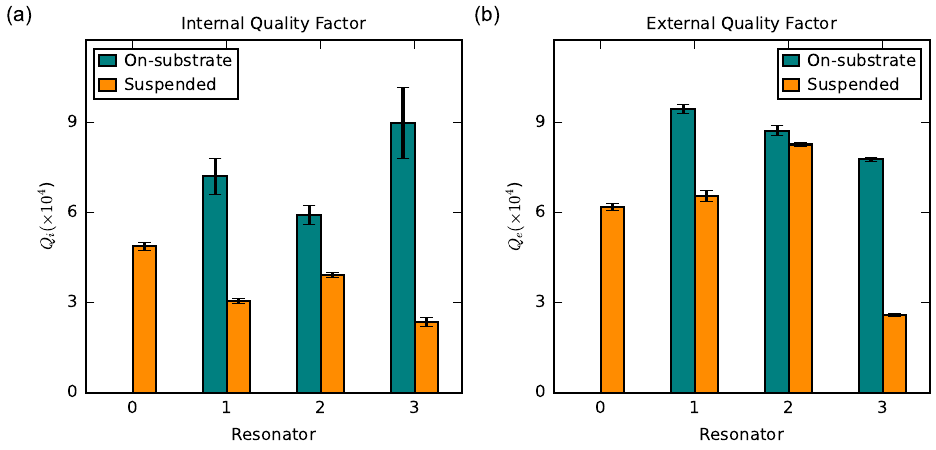}
\caption{Resonator quality factors. (a) $Q_i$ and (b) $Q_e$ values of the on-substrate (teal) and suspended (orange) arrays. The resonators sharing an index have the same number of junctions, where the index $[0,1,2,3]$ corresponds to junction numbers $[400, 300, 200, 100]$, respectively.}
	\label{fig:array_Qi_bar_plot}
\end{figure}

While the summarizing values were described in the main text, we present the results for each array's internal quality factor ($Q_i$) and external quality factor ($Q_e$) below, comparing on-substrate versus suspended arrays in Fig.~\ref{fig:array_Qi_bar_plot}. The resonators are ordered starting from the lowest frequency resonator with the greatest number of junctions. Each average value is the mean of the quality factor value corresponding to $\bar{n} \sim 1$ and its two immediate neighboring points, where ascending $\bar{n}$ values were taken sequentially in time. This decision is motivated by the array's instabilities in time creating some variation in the device response and consequent extracted $Q_i$ and $Q_e$ values. The error bars are the standard deviation of the three points.

From these values, we observe a consistent trend of a reduced $Q_i$ from the on-substrate to suspended arrays. There is no strong trend in $Q_i$ dependent upon the number of junctions in the array. Furthermore, from simulation and optimization iterations in the chip design, all resonator $Q_i$ and $Q_e$ values were comparable to each other.

\subsection{Individual Resonator Response with Varying Power}
\begin{figure}[h!]
\includegraphics[scale=1]{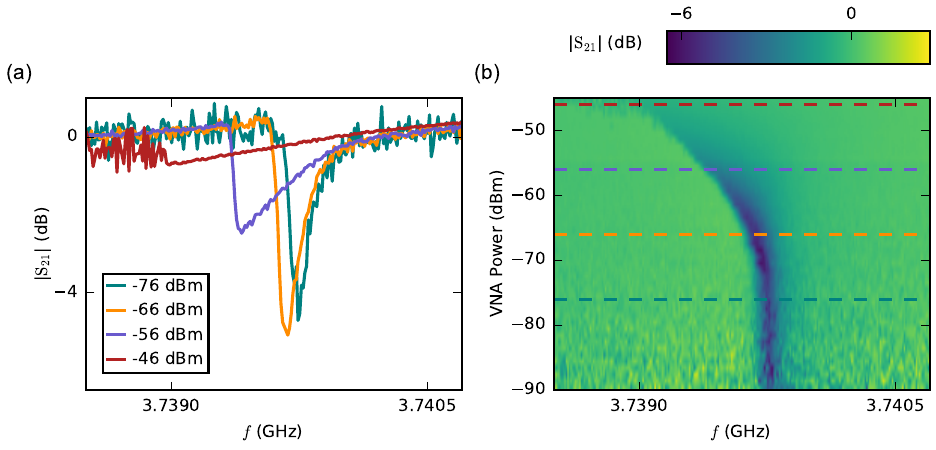}
\caption{Resonator response at varying powers. (a) Individual traces of the resonator's amplitude response, which becomes increasingly asymmetric at higher drive powers. The legend indicates the VNA drive power. (b) Corresponding power sweep from which the traces in (a) are selected from, indicated by the overlaid lines of the same colors.}
	\label{fig_supp:power_sweep_linecuts}
\end{figure}

Here, we provide further illustration of the resonator response versus power, as in Fig.~\ref{fig_supp:power_sweep_linecuts}. Panel (a) contains individual amplitude traces of an array's response at various powers, where the corresponding power from the larger sweep is designated in panel (b). This data exhibits how the array's frequency decreases with increasing drive power due to its self-Kerr coefficient, while it also becomes more asymmetric.

\subsection{Drive Line Calibration}
In the main text, we describe our experiments to extract the Kerr coefficients of the JJ arrays. To do so, we must convert the VNA drive power outside the dilution refrigerator to the average photon occupation ($\bar{n}$) of the resonator at the mixing chamber plate. We perform this conversion by estimating the drive line attenuation, beginning with measuring the line attenuation at room temperature. To compensate for changes in cable attenuation between room temperature and when the dilution refrigerator is cooled down, we shorted two nominally identical drive lines to each other at the mixing chamber plate, yielding one continuous chain in and out of the dilution refrigerator. By comparing the measured attenuation at room temperature and when cooled down, we extract the difference in cable attenuation to achieve a more calibrated attenuation of the drive line. At room temperature, the cables are 0.85 dB more attenuating than when cold. We incorporate this correction into our attenuation value used to convert from VNA drive power to $\bar{n}$.

However, we note that there still remains uncertainty in our attenuation calibration, which will cause the greatest uncertainty in our extracted Kerr coefficients. Therefore, in the main text, we include our corresponding estimates for the Kerr coefficients while acknowledging uncertainty in the numerical values. Nonetheless, the ratios of Kerr coefficients of one array to another are not affected by the exact attenuation value, so the resonators will always yield the observed trend that follows expectations. 

\subsection{Determining Ground Capacitance}
\label{supp:GND_C}

\begin{figure}[h!]
\includegraphics[]{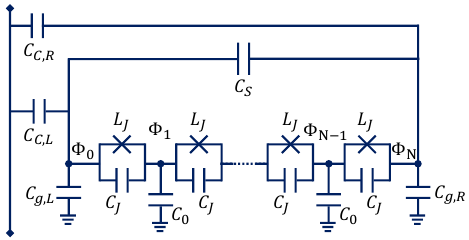}
\caption{Circuit diagram of the capacitively shunted JJ array, which also has capacitances to a transmission line for measurement.}
	\label{fig:JJ_circuit_diagram}
\end{figure}

\begingroup
\setlength{\tabcolsep}{6pt}
\renewcommand{\arraystretch}{1.2}
\begin{table}
\centering
\begin{tabular}{|c|c|c|}
 \hline
 Parameter & Value & Source \\
 \hline 
 $L_{J, \text{ Substrate}}$ & 0.91 nH & Probe results \\
 \hline 
 $L_{J, \text{ Etched}}$ & 1.10 nH & Probe results \\
 \hline 
 $\epsilon$ & $9.0 \epsilon_0$ & Ref. \cite{Weides_phase_qb_fab} \\
 \hline 
 Junction area ($A$) & 1 $\mathrm{\mu m}^2$ & SEM image \\
 \hline 
 Junction thickness ($d$) & 2.5 nm & Past TEM \cite{Kim_laser_annealing} \\
 \hline 
 $C_{J}$ & 20 fF & Above values \\
 \hline 
 $C_{S,\text{ Substrate}}$ & [0.51, 0.51, 0.53, 0.53, 0.58] fF & Ansys simulation \\
 \hline 
 $C_{S,\text{ Etched}}$ & [0.48, 0.48, 0.50, 0.50, 0.55] fF & Ansys simulation \\
 \hline 
 $C_{C,L,\text{ Substrate}}$ & [0.99, 0.94, 0.95, 0.85, 0.93] fF & Ansys simulation \\
 \hline 
 $C_{C,L,\text{ Etched}}$ & [0.99, 0.94, 0.94, 0.85, 0.93] fF & Ansys simulation \\
 \hline 
 $C_{C,R,\text{ Substrate}}$ & [0.34, 0.32, 0.31, 0.29, 0.26] fF & Ansys simulation \\
 \hline 
 $C_{C,R,\text{ Etched}}$ & [0.34, 0.32, 0.3, 0.28, 0.26] fF & Ansys simulation \\
 \hline 
 $C_{g,L,\text{ Substrate}}$ & [5.89, 5.84, 6.29, 6.37, 7.72] fF & Ansys simulation \\
 \hline 
 $C_{g,L,\text{ Etched}}$ & [5.81, 5.8, 6.26, 6.34, 7.67] fF & Ansys simulation \\
 \hline 
 $C_{g,R,\text{ Substrate}}$ & [6.55, 6.43, 6.47, 6.49, 6.41] fF & Ansys simulation \\
 \hline 
 $C_{g,R,\text{ Etched}}$ & [6.46, 6.40, 6.43, 6.46, 6.46] fF & Ansys simulation \\
 \hline 
 N & [500, 400, 300, 200, 100] & Design \\
 \hline 
\end{tabular}
\caption{Table of the parameters used to extract the ground capacitance of the arrays. $\epsilon$, the junction area, and the junction thickness are used to determine the junction capacitance, taken as a parallel plate capacitor $C_J = \epsilon A / d$. The ``Substrate" subscript indicates values for on-substrate devices, while the ``Etched" subscript indicates values for suspended devices.}
\label{table:wl_parameters}
\end{table}
\endgroup

While Eq. \ref{eqn:isolated_array_freq} in the main text describes an isolated series array of Josesphson junctions, the devices in this study are terminated by capacitor paddles, which provide a shunting capacitance across the array and also serve to couple the array to the transmission line for measurement. We illustrate the corresponding circuit diagram for our devices in Fig. \ref{fig:JJ_circuit_diagram}. For $N$ junctions, this circuit yields the Lagrangian:

\begin{equation} \label{eq:JJ_array_L}
\begin{split}
    \mathcal{L} &= \frac{1}{2}\left(C_{g,L}+C_{C,L}\right)\dot{\Phi}_0^2 + \frac{1}{2}\sum_{n=0}^{N-1}C_J\left(\dot{\Phi}_{n+1}-\dot{\Phi}_n\right)^2 + \frac{1}{2}\left(C_{g,R}+C_{C,R}\right)\dot{\Phi}_N^2 + \frac{1}{2}\sum_{n=1}^{N-1}C_0\dot{\Phi}_n^2 + \frac{1}{2}C_S\left(\dot{\Phi}_N - \dot{\Phi}_0\right)^2 \\ 
    &+ \sum_{n=0}^{N-1}E_J\cos(\phi_{n+1}-\phi_n).
\end{split}
\end{equation}
From here, we follow the same methods as in Ref. \cite{Weissl_Kerr_coefficients}, expanding the cosine potential term to second order and fitting the array frequencies. We use the ground capacitance, $C_0$, as the free parameter, and we set the remaining variables from experimental measurements, referenced results, or simulated values as indicated in Tab. \ref{table:wl_parameters}.

Although each chip has five arrays, we are able to directly measure three from the chip with on-substrate resonators, while we measure four from the chip with suspended resonators where the frequencies were relatively higher. We could not measure all arrays due to the measurement bandwidth of the setup being approximately limited to the typical $4 - 8$ GHz range. Thus, while any features a bit below 4 GHz were still visible, due to filtering, any features too far below would not be resolvable. The remaining resonators likely are below this cutoff.
We find that of the three arrays measured on both devices, the etched devices all have lower $C_0$ compared to their on-substrate counterparts. The extracted values are compiled in Tab. \ref{table:C0_vals}. We acknowledge that the fitting routine to extract $C_0$ is only based on the experimentally measured fundamental frequency instead of fitting to multiple modes, as other work has done \cite{Weissl_Kerr_coefficients}, so it relies upon many simulated and referenced values. All the same, the results still demonstrate a significant reduction in $C_0$ from the etching method.

\begingroup
\setlength{\tabcolsep}{6pt}
\renewcommand{\arraystretch}{1.2}
\begin{table}
\centering
\begin{tabular}{|c|c|c|c|}
 \hline
 Junction Num. (N) & On-substrate $C_0$ (aF) & Etched $C_0$ (aF) & Relative Change in $C_0$ by Etching \\
 \hline 
 400 & N/A & 15 & N/A \\
 \hline 
 300 & 118 & 14 & -0.88 \\
 \hline 
 200 & 158 & 60 & -0.62 \\
 \hline 
 100 & 293 & 83 & -0.72 \\
 \hline 
\end{tabular}
\caption{Extracted ground capacitance ($C_0$) values, which are in the approximate range of values from other work \cite{Masluk_JJ_array, Weissl_Kerr_coefficients}. Note that the values decrease from the on-substrate to etched variants. This is encapsulated in the final column of the ``Relative Change by Etching," or the difference between the etched and on-substrate capacitance values relative to the on-substrate value.}
\label{table:C0_vals}
\end{table}
\endgroup

\subsection{Qubit Wafer Scale Room Temperature Statistics}

Here, we provide statistics on the room-temperature probing of the qubit chips, summarized in Tab. \ref{table:qubit_probing}. For each fluxonium or suspended fluxonium, we perform two-point probing as was done for the arrays. We partition the results into multiple categories, analogous to the qubit conditions described in the main text. We summarize results of 3 chips that are conventionally cleaned (``On-substrate, non-etched chip," similar to Device A in the main text), and 7 chips that are etched, where qubits are either all on-substrate (``On-substrate, etched chip," like Device B) or where the array is suspended (``Suspended, etched chip," like Device C). We present the resistance value mean and standard deviation (Std. Dev.). Due to the robustness and low variability of the arrays, we attribute the variability in the fluxonium devices primarily to the small junction. We find that all conditions have similar variability.

\begingroup
\setlength{\tabcolsep}{6pt}
\renewcommand{\arraystretch}{1.2}
\begin{table}
\centering
\begin{tabular}{|c|c|c|c|c|}
 \hline
 Condition & Num. Qubits & Mean (k$\Omega$) & Std. Dev. (k$\Omega$) & Std. Dev./Mean \\
 \hline 
 On-substrate, non-etched chip & 24 & 55.3 & 6.4 & 0.12 \\
 \hline 
 On-substrate, etched chip & 28 & 47.6 & 4.5 & 0.10\\
 \hline 
 Total on-substrate, both chips & 52 & 52.6 & 6.9 & 0.13 \\
 \hline
 Suspended, etched chip & 28 & 50.8 & 5.3 & 0.10 \\
 \hline 
\end{tabular}
\caption{Results from room-temperature probing of qubits. The rows, in descending order, correspond to: fluxonium qubits on non-etched chips, fluxonium qubits on etched chips, all fluxonium qubits on non-etched and etched chips, and lastly suspended fluxonium qubits.}
\label{table:qubit_probing}
\end{table}
\endgroup

\subsection{Device B Spectrum}
\begin{figure}[h!]
\includegraphics[scale=1]{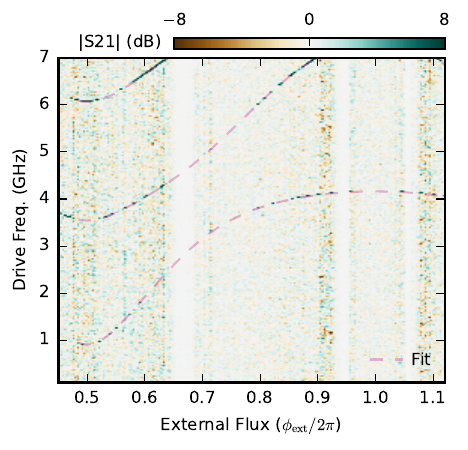}
\caption{Device B's transition frequency spectrum. Extracted $0-1$, $0-2$, and $0-3$ transitions are overlaid in magenta.}
	\label{fig:Device_B_spectrum}
\end{figure}

Here in Fig. \ref{fig:Device_B_spectrum}, we include the transition frequency spectrum of Device B, the on-substrate fluxonium on the same chip as the etched suspended fluxonium, Device C.

\subsection{Thermal Photon Dephasing in Device C}

\begin{figure}[h!]
\includegraphics[scale=0.8]{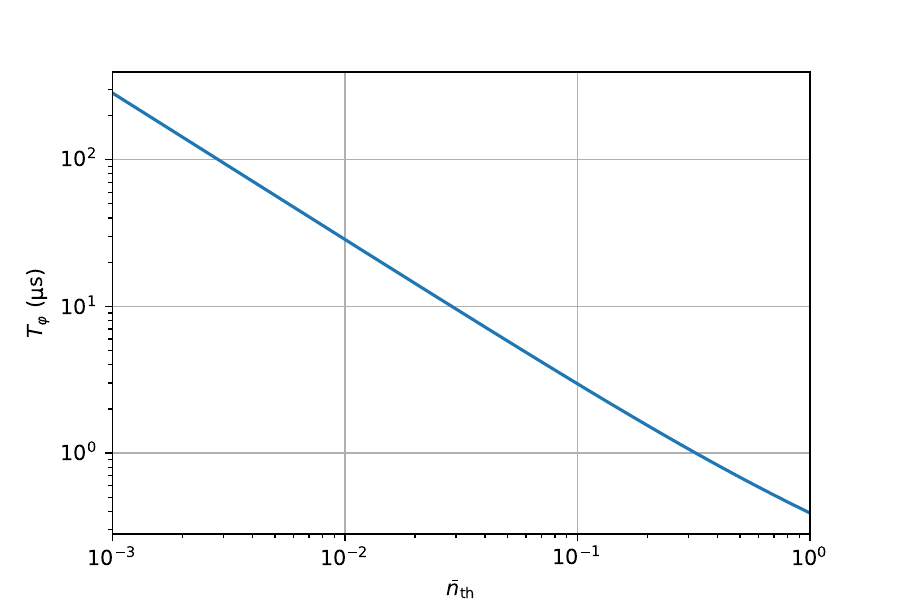}
\caption{Dephasing times $T_\varphi$ originating from the fluctuation of resonator photons $\bar{n}_\mathrm{th}$ in Device C.}
	\label{fig:thermal_dephasing}
\end{figure}

To account for the reduced coherence times measured in Device C, we can investigate the expected limits imposed by thermal photon dephasing. The qubit is coupled with a strength $g\sim 100~\mathrm{MHz}$ to a resonator of frequency $\omega_R/2\pi = 7.18~\mathrm{GHz}$ and linewidth $\kappa_R/2\pi = 0.8~\mathrm{MHz}$. Using the extracted parameters $[E_J, E_C, E_L] = [2.59, 1.01, 0.42]$ GHz to find all relevant energy transitions, we calculate a dispersive shift of $|\chi_{01}|/2\pi = 1.38~\mathrm{MHz}$ using~\cite{zhu2013circuit,Nguyen_blueprint_fluxonium}
\begin{equation}
        \chi_{01} = g^2 \left[\sum_{l\neq 0} |n_{0l}|^2 \frac{2\omega_{0l}}{\omega_{0l}^2 - \omega_R^2} - \sum_{l\neq 1}|n_{1l}|^2\frac{2\omega_{1l}}{\omega_{1l}^2 - \omega_R^2}  \right].
        \label{eqn:dispersive_shift_2nd}
\end{equation}
From here, we can predict the thermal photon dephasing rate $\Gamma_\mathrm{\varphi}$~\cite{rigetti2012superconducting}
\begin{equation}
    \Gamma_\mathrm{\varphi} = \frac{\kappa_R}{2}\mathrm{Re}\left[\sqrt{\left(1+\frac{i\chi_{01}}{\kappa_R} \right)^2 + \frac{4i\chi_{01}\mathrm{n^{eff}_{th}}}{\kappa_R}}-1 \right].
\end{equation}
Sweeping $\mathrm{\bar{n}_{th}}$, we plot the corresponding dephasing time $T_\varphi = 1/\Gamma_\mathrm{\varphi}$ in Fig.~\ref{fig:thermal_dephasing}.

Typical thermal photon numbers in planar superconducting devices are in the range of $n_\mathrm{th}\sim 10^{-2}$~\cite{yan2018distinguishing}, which would limit the dephasing times to a few tens of microseconds. From the coherence statistics of Device C, we find an average Ramsey dephasing time $T_2^*=20\pm 3~\mathrm{\mu s}$ and an average echo dephasing time $T_\mathrm{2,Echo}=26\pm 6~\mathrm{\mu s}$. These similar values further support the hypothesis that the device's coherence time is limited by thermal photon dephasing. 

\begin{thebibliography}{68}%
\makeatletter
\providecommand \@ifxundefined [1]{%
 \@ifx{#1\undefined}
}%
\providecommand \@ifnum [1]{%
 \ifnum #1\expandafter \@firstoftwo
 \else \expandafter \@secondoftwo
 \fi
}%
\providecommand \@ifx [1]{%
 \ifx #1\expandafter \@firstoftwo
 \else \expandafter \@secondoftwo
 \fi
}%
\providecommand \natexlab [1]{#1}%
\providecommand \enquote  [1]{``#1''}%
\providecommand \bibnamefont  [1]{#1}%
\providecommand \bibfnamefont [1]{#1}%
\providecommand \citenamefont [1]{#1}%
\providecommand \href@noop [0]{\@secondoftwo}%
\providecommand \href [0]{\begingroup \@sanitize@url \@href}%
\providecommand \@href[1]{\@@startlink{#1}\@@href}%
\providecommand \@@href[1]{\endgroup#1\@@endlink}%
\providecommand \@sanitize@url [0]{\catcode `\\12\catcode `\$12\catcode `\&12\catcode `\#12\catcode `\^12\catcode `\_12\catcode `\%12\relax}%
\providecommand \@@startlink[1]{}%
\providecommand \@@endlink[0]{}%
\providecommand \url  [0]{\begingroup\@sanitize@url \@url }%
\providecommand \@url [1]{\endgroup\@href {#1}{\urlprefix }}%
\providecommand \urlprefix  [0]{URL }%
\providecommand \Eprint [0]{\href }%
\providecommand \doibase [0]{https://doi.org/}%
\providecommand \selectlanguage [0]{\@gobble}%
\providecommand \bibinfo  [0]{\@secondoftwo}%
\providecommand \bibfield  [0]{\@secondoftwo}%
\providecommand \translation [1]{[#1]}%
\providecommand \BibitemOpen [0]{}%
\providecommand \bibitemStop [0]{}%
\providecommand \bibitemNoStop [0]{.\EOS\space}%
\providecommand \EOS [0]{\spacefactor3000\relax}%
\providecommand \BibitemShut  [1]{\csname bibitem#1\endcsname}%
\let\auto@bib@innerbib\@empty
\bibitem [{\citenamefont {Manucharyan}\ \emph {et~al.}(2009)\citenamefont {Manucharyan}, \citenamefont {Koch}, \citenamefont {Glazman},\ and\ \citenamefont {Devoret}}]{Manucharyan_fluxonium}%
  \BibitemOpen
  \bibfield  {author} {\bibinfo {author} {\bibfnamefont {V.~E.}\ \bibnamefont {Manucharyan}}, \bibinfo {author} {\bibfnamefont {J.}~\bibnamefont {Koch}}, \bibinfo {author} {\bibfnamefont {L.~I.}\ \bibnamefont {Glazman}},\ and\ \bibinfo {author} {\bibfnamefont {M.~H.}\ \bibnamefont {Devoret}},\ }\bibfield  {title} {\bibinfo {title} {Fluxonium: Single cooper-pair circuit free of charge offsets},\ }\href {https://doi.org/10.1126/science.1175552} {\bibfield  {journal} {\bibinfo  {journal} {Science}\ }\textbf {\bibinfo {volume} {326}},\ \bibinfo {pages} {113–116} (\bibinfo {year} {2009})}\BibitemShut {NoStop}%
\bibitem [{\citenamefont {Pop}\ \emph {et~al.}(2014)\citenamefont {Pop}, \citenamefont {Geerlings}, \citenamefont {Catelani}, \citenamefont {Schoelkopf}, \citenamefont {Glazman},\ and\ \citenamefont {Devoret}}]{pop2014coherent}%
  \BibitemOpen
  \bibfield  {author} {\bibinfo {author} {\bibfnamefont {I.~M.}\ \bibnamefont {Pop}}, \bibinfo {author} {\bibfnamefont {K.}~\bibnamefont {Geerlings}}, \bibinfo {author} {\bibfnamefont {G.}~\bibnamefont {Catelani}}, \bibinfo {author} {\bibfnamefont {R.~J.}\ \bibnamefont {Schoelkopf}}, \bibinfo {author} {\bibfnamefont {L.~I.}\ \bibnamefont {Glazman}},\ and\ \bibinfo {author} {\bibfnamefont {M.~H.}\ \bibnamefont {Devoret}},\ }\bibfield  {title} {\bibinfo {title} {Coherent suppression of electromagnetic dissipation due to superconducting quasiparticles},\ }\href@noop {} {\bibfield  {journal} {\bibinfo  {journal} {Nature}\ }\textbf {\bibinfo {volume} {508}},\ \bibinfo {pages} {369} (\bibinfo {year} {2014})}\BibitemShut {NoStop}%
\bibitem [{\citenamefont {Nguyen}\ \emph {et~al.}(2019)\citenamefont {Nguyen}, \citenamefont {Lin}, \citenamefont {Somoroff}, \citenamefont {Mencia}, \citenamefont {Grabon},\ and\ \citenamefont {Manucharyan}}]{Nguyen_high_coherence_fluxonium}%
  \BibitemOpen
  \bibfield  {author} {\bibinfo {author} {\bibfnamefont {L.~B.}\ \bibnamefont {Nguyen}}, \bibinfo {author} {\bibfnamefont {Y.-H.}\ \bibnamefont {Lin}}, \bibinfo {author} {\bibfnamefont {A.}~\bibnamefont {Somoroff}}, \bibinfo {author} {\bibfnamefont {R.}~\bibnamefont {Mencia}}, \bibinfo {author} {\bibfnamefont {N.}~\bibnamefont {Grabon}},\ and\ \bibinfo {author} {\bibfnamefont {V.~E.}\ \bibnamefont {Manucharyan}},\ }\bibfield  {title} {\bibinfo {title} {High-coherence fluxonium qubit},\ }\href {https://doi.org/10.1103/PhysRevX.9.041041} {\bibfield  {journal} {\bibinfo  {journal} {Phys. Rev. X}\ }\textbf {\bibinfo {volume} {9}},\ \bibinfo {pages} {041041} (\bibinfo {year} {2019})}\BibitemShut {NoStop}%
\bibitem [{\citenamefont {Zhang}\ \emph {et~al.}(2021)\citenamefont {Zhang}, \citenamefont {Chakram}, \citenamefont {Roy}, \citenamefont {Earnest}, \citenamefont {Lu}, \citenamefont {Huang}, \citenamefont {Weiss}, \citenamefont {Koch},\ and\ \citenamefont {Schuster}}]{zhang2021universal}%
  \BibitemOpen
  \bibfield  {author} {\bibinfo {author} {\bibfnamefont {H.}~\bibnamefont {Zhang}}, \bibinfo {author} {\bibfnamefont {S.}~\bibnamefont {Chakram}}, \bibinfo {author} {\bibfnamefont {T.}~\bibnamefont {Roy}}, \bibinfo {author} {\bibfnamefont {N.}~\bibnamefont {Earnest}}, \bibinfo {author} {\bibfnamefont {Y.}~\bibnamefont {Lu}}, \bibinfo {author} {\bibfnamefont {Z.}~\bibnamefont {Huang}}, \bibinfo {author} {\bibfnamefont {D.~K.}\ \bibnamefont {Weiss}}, \bibinfo {author} {\bibfnamefont {J.}~\bibnamefont {Koch}},\ and\ \bibinfo {author} {\bibfnamefont {D.~I.}\ \bibnamefont {Schuster}},\ }\bibfield  {title} {\bibinfo {title} {Universal fast-flux control of a coherent, low-frequency qubit},\ }\href {https://doi.org/10.1103/PhysRevX.11.011010} {\bibfield  {journal} {\bibinfo  {journal} {Phys. Rev. X}\ }\textbf {\bibinfo {volume} {11}},\ \bibinfo {pages} {011010} (\bibinfo {year} {2021})}\BibitemShut {NoStop}%
\bibitem [{\citenamefont {Brooks}\ \emph {et~al.}(2013)\citenamefont {Brooks}, \citenamefont {Kitaev},\ and\ \citenamefont {Preskill}}]{brooks2013protected}%
  \BibitemOpen
  \bibfield  {author} {\bibinfo {author} {\bibfnamefont {P.}~\bibnamefont {Brooks}}, \bibinfo {author} {\bibfnamefont {A.}~\bibnamefont {Kitaev}},\ and\ \bibinfo {author} {\bibfnamefont {J.}~\bibnamefont {Preskill}},\ }\bibfield  {title} {\bibinfo {title} {Protected gates for superconducting qubits},\ }\href {https://doi.org/10.1103/PhysRevA.87.052306} {\bibfield  {journal} {\bibinfo  {journal} {Phys. Rev. A}\ }\textbf {\bibinfo {volume} {87}},\ \bibinfo {pages} {052306} (\bibinfo {year} {2013})}\BibitemShut {NoStop}%
\bibitem [{\citenamefont {Gyenis}\ \emph {et~al.}(2021)\citenamefont {Gyenis}, \citenamefont {Di~Paolo}, \citenamefont {Koch}, \citenamefont {Blais}, \citenamefont {Houck},\ and\ \citenamefont {Schuster}}]{Gyenis2021}%
  \BibitemOpen
  \bibfield  {author} {\bibinfo {author} {\bibfnamefont {A.}~\bibnamefont {Gyenis}}, \bibinfo {author} {\bibfnamefont {A.}~\bibnamefont {Di~Paolo}}, \bibinfo {author} {\bibfnamefont {J.}~\bibnamefont {Koch}}, \bibinfo {author} {\bibfnamefont {A.}~\bibnamefont {Blais}}, \bibinfo {author} {\bibfnamefont {A.~A.}\ \bibnamefont {Houck}},\ and\ \bibinfo {author} {\bibfnamefont {D.~I.}\ \bibnamefont {Schuster}},\ }\bibfield  {title} {\bibinfo {title} {Moving beyond the transmon: Noise-protected superconducting quantum circuits},\ }\href {https://doi.org/10.1103/prxquantum.2.030101} {\bibfield  {journal} {\bibinfo  {journal} {PRX Quantum}\ }\textbf {\bibinfo {volume} {2}},\ \bibinfo {pages} {030101} (\bibinfo {year} {2021})}\BibitemShut {NoStop}%
\bibitem [{\citenamefont {Kalashnikov}\ \emph {et~al.}(2020)\citenamefont {Kalashnikov}, \citenamefont {Hsieh}, \citenamefont {Zhang}, \citenamefont {Lu}, \citenamefont {Kamenov}, \citenamefont {Di~Paolo}, \citenamefont {Blais}, \citenamefont {Gershenson},\ and\ \citenamefont {Bell}}]{Kalashnikov_bifluxon}%
  \BibitemOpen
  \bibfield  {author} {\bibinfo {author} {\bibfnamefont {K.}~\bibnamefont {Kalashnikov}}, \bibinfo {author} {\bibfnamefont {W.~T.}\ \bibnamefont {Hsieh}}, \bibinfo {author} {\bibfnamefont {W.}~\bibnamefont {Zhang}}, \bibinfo {author} {\bibfnamefont {W.-S.}\ \bibnamefont {Lu}}, \bibinfo {author} {\bibfnamefont {P.}~\bibnamefont {Kamenov}}, \bibinfo {author} {\bibfnamefont {A.}~\bibnamefont {Di~Paolo}}, \bibinfo {author} {\bibfnamefont {A.}~\bibnamefont {Blais}}, \bibinfo {author} {\bibfnamefont {M.~E.}\ \bibnamefont {Gershenson}},\ and\ \bibinfo {author} {\bibfnamefont {M.}~\bibnamefont {Bell}},\ }\bibfield  {title} {\bibinfo {title} {Bifluxon: Fluxon-parity-protected superconducting qubit},\ }\href {https://doi.org/10.1103/PRXQuantum.1.010307} {\bibfield  {journal} {\bibinfo  {journal} {PRX Quantum}\ }\textbf {\bibinfo {volume} {1}},\ \bibinfo {pages} {010307} (\bibinfo {year} {2020})}\BibitemShut {NoStop}%
\bibitem [{\citenamefont {Bell}\ \emph {et~al.}(2014)\citenamefont {Bell}, \citenamefont {Paramanandam}, \citenamefont {Ioffe},\ and\ \citenamefont {Gershenson}}]{bell2014protected}%
  \BibitemOpen
  \bibfield  {author} {\bibinfo {author} {\bibfnamefont {M.~T.}\ \bibnamefont {Bell}}, \bibinfo {author} {\bibfnamefont {J.}~\bibnamefont {Paramanandam}}, \bibinfo {author} {\bibfnamefont {L.~B.}\ \bibnamefont {Ioffe}},\ and\ \bibinfo {author} {\bibfnamefont {M.~E.}\ \bibnamefont {Gershenson}},\ }\bibfield  {title} {\bibinfo {title} {Protected {Josephson} rhombus chains},\ }\href {https://doi.org/10.1103/PhysRevLett.112.167001} {\bibfield  {journal} {\bibinfo  {journal} {Phys. Rev. Lett.}\ }\textbf {\bibinfo {volume} {112}},\ \bibinfo {pages} {167001} (\bibinfo {year} {2014})}\BibitemShut {NoStop}%
\bibitem [{\citenamefont {Larsen}\ \emph {et~al.}(2020)\citenamefont {Larsen}, \citenamefont {Gershenson}, \citenamefont {Casparis}, \citenamefont {Kringh\o{}j}, \citenamefont {Pearson}, \citenamefont {McNeil}, \citenamefont {Kuemmeth}, \citenamefont {Krogstrup}, \citenamefont {Petersson},\ and\ \citenamefont {Marcus}}]{larsen2020protected}%
  \BibitemOpen
  \bibfield  {author} {\bibinfo {author} {\bibfnamefont {T.~W.}\ \bibnamefont {Larsen}}, \bibinfo {author} {\bibfnamefont {M.~E.}\ \bibnamefont {Gershenson}}, \bibinfo {author} {\bibfnamefont {L.}~\bibnamefont {Casparis}}, \bibinfo {author} {\bibfnamefont {A.}~\bibnamefont {Kringh\o{}j}}, \bibinfo {author} {\bibfnamefont {N.~J.}\ \bibnamefont {Pearson}}, \bibinfo {author} {\bibfnamefont {R.~P.~G.}\ \bibnamefont {McNeil}}, \bibinfo {author} {\bibfnamefont {F.}~\bibnamefont {Kuemmeth}}, \bibinfo {author} {\bibfnamefont {P.}~\bibnamefont {Krogstrup}}, \bibinfo {author} {\bibfnamefont {K.~D.}\ \bibnamefont {Petersson}},\ and\ \bibinfo {author} {\bibfnamefont {C.~M.}\ \bibnamefont {Marcus}},\ }\bibfield  {title} {\bibinfo {title} {Parity-protected superconductor-semiconductor qubit},\ }\href {https://doi.org/10.1103/PhysRevLett.125.056801} {\bibfield  {journal} {\bibinfo  {journal} {Phys. Rev. Lett.}\ }\textbf {\bibinfo {volume} {125}},\ \bibinfo {pages} {056801} (\bibinfo {year} {2020})}\BibitemShut {NoStop}%
\bibitem [{\citenamefont {Smith}\ \emph {et~al.}(2022)\citenamefont {Smith}, \citenamefont {Villiers}, \citenamefont {Marquet}, \citenamefont {Palomo}, \citenamefont {Delbecq}, \citenamefont {Kontos}, \citenamefont {Campagne-Ibarcq}, \citenamefont {Dou\ifmmode~\mbox{\c{c}}\else \c{c}\fi{}ot},\ and\ \citenamefont {Leghtas}}]{smith2022magnifying}%
  \BibitemOpen
  \bibfield  {author} {\bibinfo {author} {\bibfnamefont {W.~C.}\ \bibnamefont {Smith}}, \bibinfo {author} {\bibfnamefont {M.}~\bibnamefont {Villiers}}, \bibinfo {author} {\bibfnamefont {A.}~\bibnamefont {Marquet}}, \bibinfo {author} {\bibfnamefont {J.}~\bibnamefont {Palomo}}, \bibinfo {author} {\bibfnamefont {M.~R.}\ \bibnamefont {Delbecq}}, \bibinfo {author} {\bibfnamefont {T.}~\bibnamefont {Kontos}}, \bibinfo {author} {\bibfnamefont {P.}~\bibnamefont {Campagne-Ibarcq}}, \bibinfo {author} {\bibfnamefont {B.}~\bibnamefont {Dou\ifmmode~\mbox{\c{c}}\else \c{c}\fi{}ot}},\ and\ \bibinfo {author} {\bibfnamefont {Z.}~\bibnamefont {Leghtas}},\ }\bibfield  {title} {\bibinfo {title} {Magnifying quantum phase fluctuations with cooper-pair pairing},\ }\href {https://doi.org/10.1103/PhysRevX.12.021002} {\bibfield  {journal} {\bibinfo  {journal} {Phys. Rev. X}\ }\textbf {\bibinfo {volume} {12}},\ \bibinfo {pages} {021002} (\bibinfo {year} {2022})}\BibitemShut {NoStop}%
\bibitem [{\citenamefont {Dodge}\ \emph {et~al.}(2023)\citenamefont {Dodge}, \citenamefont {Liu}, \citenamefont {Klots}, \citenamefont {Cole}, \citenamefont {Shearrow}, \citenamefont {Senatore}, \citenamefont {Zhu}, \citenamefont {Ioffe}, \citenamefont {McDermott},\ and\ \citenamefont {Plourde}}]{dodge2023hardware}%
  \BibitemOpen
  \bibfield  {author} {\bibinfo {author} {\bibfnamefont {K.}~\bibnamefont {Dodge}}, \bibinfo {author} {\bibfnamefont {Y.}~\bibnamefont {Liu}}, \bibinfo {author} {\bibfnamefont {A.~R.}\ \bibnamefont {Klots}}, \bibinfo {author} {\bibfnamefont {B.}~\bibnamefont {Cole}}, \bibinfo {author} {\bibfnamefont {A.}~\bibnamefont {Shearrow}}, \bibinfo {author} {\bibfnamefont {M.}~\bibnamefont {Senatore}}, \bibinfo {author} {\bibfnamefont {S.}~\bibnamefont {Zhu}}, \bibinfo {author} {\bibfnamefont {L.~B.}\ \bibnamefont {Ioffe}}, \bibinfo {author} {\bibfnamefont {R.}~\bibnamefont {McDermott}},\ and\ \bibinfo {author} {\bibfnamefont {B.~L.~T.}\ \bibnamefont {Plourde}},\ }\bibfield  {title} {\bibinfo {title} {Hardware implementation of quantum stabilizers in superconducting circuits},\ }\href {https://doi.org/10.1103/PhysRevLett.131.150602} {\bibfield  {journal} {\bibinfo  {journal} {Phys. Rev. Lett.}\ }\textbf {\bibinfo {volume} {131}},\ \bibinfo {pages} {150602} (\bibinfo {year} {2023})}\BibitemShut {NoStop}%
\bibitem [{\citenamefont {Pechenezhskiy}\ \emph {et~al.}(2020)\citenamefont {Pechenezhskiy}, \citenamefont {Mencia}, \citenamefont {Nguyen}, \citenamefont {Lin},\ and\ \citenamefont {Manucharyan}}]{Pechenezhskiy_blochnium}%
  \BibitemOpen
  \bibfield  {author} {\bibinfo {author} {\bibfnamefont {I.~V.}\ \bibnamefont {Pechenezhskiy}}, \bibinfo {author} {\bibfnamefont {R.~A.}\ \bibnamefont {Mencia}}, \bibinfo {author} {\bibfnamefont {L.~B.}\ \bibnamefont {Nguyen}}, \bibinfo {author} {\bibfnamefont {Y.-H.}\ \bibnamefont {Lin}},\ and\ \bibinfo {author} {\bibfnamefont {V.~E.}\ \bibnamefont {Manucharyan}},\ }\bibfield  {title} {\bibinfo {title} {The superconducting quasicharge qubit},\ }\href {https://doi.org/10.1038/s41586-020-2687-9} {\bibfield  {journal} {\bibinfo  {journal} {Nature}\ }\textbf {\bibinfo {volume} {585}},\ \bibinfo {pages} {368–371} (\bibinfo {year} {2020})}\BibitemShut {NoStop}%
\bibitem [{\citenamefont {Maleeva}\ \emph {et~al.}(2018)\citenamefont {Maleeva}, \citenamefont {Grünhaupt}, \citenamefont {Klein}, \citenamefont {Levy-Bertrand}, \citenamefont {Dupre}, \citenamefont {Calvo}, \citenamefont {Valenti}, \citenamefont {Winkel}, \citenamefont {Friedrich}, \citenamefont {Wernsdorfer}, \citenamefont {Ustinov}, \citenamefont {Rotzinger}, \citenamefont {Monfardini}, \citenamefont {Fistul},\ and\ \citenamefont {Pop}}]{Maleeva_CQED_grAl}%
  \BibitemOpen
  \bibfield  {author} {\bibinfo {author} {\bibfnamefont {N.}~\bibnamefont {Maleeva}}, \bibinfo {author} {\bibfnamefont {L.}~\bibnamefont {Grünhaupt}}, \bibinfo {author} {\bibfnamefont {T.}~\bibnamefont {Klein}}, \bibinfo {author} {\bibfnamefont {F.}~\bibnamefont {Levy-Bertrand}}, \bibinfo {author} {\bibfnamefont {O.}~\bibnamefont {Dupre}}, \bibinfo {author} {\bibfnamefont {M.}~\bibnamefont {Calvo}}, \bibinfo {author} {\bibfnamefont {F.}~\bibnamefont {Valenti}}, \bibinfo {author} {\bibfnamefont {P.}~\bibnamefont {Winkel}}, \bibinfo {author} {\bibfnamefont {F.}~\bibnamefont {Friedrich}}, \bibinfo {author} {\bibfnamefont {W.}~\bibnamefont {Wernsdorfer}}, \bibinfo {author} {\bibfnamefont {A.~V.}\ \bibnamefont {Ustinov}}, \bibinfo {author} {\bibfnamefont {H.}~\bibnamefont {Rotzinger}}, \bibinfo {author} {\bibfnamefont {A.}~\bibnamefont {Monfardini}}, \bibinfo {author} {\bibfnamefont {M.~V.}\ \bibnamefont {Fistul}},\ and\ \bibinfo {author} {\bibfnamefont {I.~M.}\ \bibnamefont {Pop}},\ }\bibfield  {title} {\bibinfo
  {title} {Circuit quantum electrodynamics of granular aluminum resonators},\ }\href {https://doi.org/10.1038/s41467-018-06386-9} {\bibfield  {journal} {\bibinfo  {journal} {Nat Commun}\ }\textbf {\bibinfo {volume} {9}},\ \bibinfo {pages} {3889} (\bibinfo {year} {2018})}\BibitemShut {NoStop}%
\bibitem [{\citenamefont {Grünhaupt}\ \emph {et~al.}(2019)\citenamefont {Grünhaupt}, \citenamefont {Spiecker}, \citenamefont {Gusenkova}, \citenamefont {Maleeva}, \citenamefont {Skacel}, \citenamefont {Takmakov}, \citenamefont {Valenti}, \citenamefont {Winkel}, \citenamefont {Rotzinger}, \citenamefont {Wernsdorfer}, \citenamefont {Ustinov},\ and\ \citenamefont {Pop}}]{Gruenhaupt2019}%
  \BibitemOpen
  \bibfield  {author} {\bibinfo {author} {\bibfnamefont {L.}~\bibnamefont {Grünhaupt}}, \bibinfo {author} {\bibfnamefont {M.}~\bibnamefont {Spiecker}}, \bibinfo {author} {\bibfnamefont {D.}~\bibnamefont {Gusenkova}}, \bibinfo {author} {\bibfnamefont {N.}~\bibnamefont {Maleeva}}, \bibinfo {author} {\bibfnamefont {S.~T.}\ \bibnamefont {Skacel}}, \bibinfo {author} {\bibfnamefont {I.}~\bibnamefont {Takmakov}}, \bibinfo {author} {\bibfnamefont {F.}~\bibnamefont {Valenti}}, \bibinfo {author} {\bibfnamefont {P.}~\bibnamefont {Winkel}}, \bibinfo {author} {\bibfnamefont {H.}~\bibnamefont {Rotzinger}}, \bibinfo {author} {\bibfnamefont {W.}~\bibnamefont {Wernsdorfer}}, \bibinfo {author} {\bibfnamefont {A.~V.}\ \bibnamefont {Ustinov}},\ and\ \bibinfo {author} {\bibfnamefont {I.~M.}\ \bibnamefont {Pop}},\ }\bibfield  {title} {\bibinfo {title} {Granular aluminium as a superconducting material for high-impedance quantum circuits},\ }\href {https://doi.org/10.1038/s41563-019-0350-3} {\bibfield  {journal} {\bibinfo
  {journal} {Nature Materials}\ }\textbf {\bibinfo {volume} {18}},\ \bibinfo {pages} {816} (\bibinfo {year} {2019})}\BibitemShut {NoStop}%
\bibitem [{\citenamefont {Kalacheva}\ \emph {et~al.}(2020)\citenamefont {Kalacheva}, \citenamefont {Fedorov}, \citenamefont {Kulakova}, \citenamefont {Zotova}, \citenamefont {Korostylev}, \citenamefont {Khrapach}, \citenamefont {Ustinov},\ and\ \citenamefont {Astafiev}}]{Kalacheva_Qi_SC_res_treatment}%
  \BibitemOpen
  \bibfield  {author} {\bibinfo {author} {\bibfnamefont {D.}~\bibnamefont {Kalacheva}}, \bibinfo {author} {\bibfnamefont {G.}~\bibnamefont {Fedorov}}, \bibinfo {author} {\bibfnamefont {A.}~\bibnamefont {Kulakova}}, \bibinfo {author} {\bibfnamefont {J.}~\bibnamefont {Zotova}}, \bibinfo {author} {\bibfnamefont {E.}~\bibnamefont {Korostylev}}, \bibinfo {author} {\bibfnamefont {I.}~\bibnamefont {Khrapach}}, \bibinfo {author} {\bibfnamefont {A.~V.}\ \bibnamefont {Ustinov}},\ and\ \bibinfo {author} {\bibfnamefont {O.~V.}\ \bibnamefont {Astafiev}},\ }\bibfield  {title} {\bibinfo {title} {Improving the quality factor of superconducting resonators by post-process surface treatment},\ }\href {https://doi.org/10.1063/5.0011900} {\bibfield  {journal} {\bibinfo  {journal} {AIP Conf. Proc.}\ }\textbf {\bibinfo {volume} {2241}},\ \bibinfo {pages} {020018} (\bibinfo {year} {2020})}\BibitemShut {NoStop}%
\bibitem [{\citenamefont {Frasca}\ \emph {et~al.}(2023)\citenamefont {Frasca}, \citenamefont {Arabadzhiev}, \citenamefont {de~Puechredon}, \citenamefont {Oppliger}, \citenamefont {Jouanny}, \citenamefont {Musio}, \citenamefont {Scigliuzzo}, \citenamefont {Minganti}, \citenamefont {Scarlino},\ and\ \citenamefont {Charbon}}]{franca2023nbn}%
  \BibitemOpen
  \bibfield  {author} {\bibinfo {author} {\bibfnamefont {S.}~\bibnamefont {Frasca}}, \bibinfo {author} {\bibfnamefont {I.}~\bibnamefont {Arabadzhiev}}, \bibinfo {author} {\bibfnamefont {S.~B.}\ \bibnamefont {de~Puechredon}}, \bibinfo {author} {\bibfnamefont {F.}~\bibnamefont {Oppliger}}, \bibinfo {author} {\bibfnamefont {V.}~\bibnamefont {Jouanny}}, \bibinfo {author} {\bibfnamefont {R.}~\bibnamefont {Musio}}, \bibinfo {author} {\bibfnamefont {M.}~\bibnamefont {Scigliuzzo}}, \bibinfo {author} {\bibfnamefont {F.}~\bibnamefont {Minganti}}, \bibinfo {author} {\bibfnamefont {P.}~\bibnamefont {Scarlino}},\ and\ \bibinfo {author} {\bibfnamefont {E.}~\bibnamefont {Charbon}},\ }\bibfield  {title} {\bibinfo {title} {Nbn films with high kinetic inductance for high-quality compact superconducting resonators},\ }\href {https://doi.org/10.1103/PhysRevApplied.20.044021} {\bibfield  {journal} {\bibinfo  {journal} {Phys. Rev. Appl.}\ }\textbf {\bibinfo {volume} {20}},\ \bibinfo {pages} {044021} (\bibinfo {year}
  {2023})}\BibitemShut {NoStop}%
\bibitem [{\citenamefont {Koolstra}\ \emph {et~al.}(2024)\citenamefont {Koolstra}, \citenamefont {Glen}, \citenamefont {Beysengulov}, \citenamefont {Byeon}, \citenamefont {Castoria}, \citenamefont {Sammon}, \citenamefont {Dizdar}, \citenamefont {Wang}, \citenamefont {Schuster}, \citenamefont {Lyon}, \citenamefont {Pollanen},\ and\ \citenamefont {Rees}}]{Koolstra_e_He_2024}%
  \BibitemOpen
  \bibfield  {author} {\bibinfo {author} {\bibfnamefont {G.}~\bibnamefont {Koolstra}}, \bibinfo {author} {\bibfnamefont {E.~O.}\ \bibnamefont {Glen}}, \bibinfo {author} {\bibfnamefont {N.~R.}\ \bibnamefont {Beysengulov}}, \bibinfo {author} {\bibfnamefont {H.}~\bibnamefont {Byeon}}, \bibinfo {author} {\bibfnamefont {K.~E.}\ \bibnamefont {Castoria}}, \bibinfo {author} {\bibfnamefont {M.}~\bibnamefont {Sammon}}, \bibinfo {author} {\bibfnamefont {B.}~\bibnamefont {Dizdar}}, \bibinfo {author} {\bibfnamefont {C.~S.}\ \bibnamefont {Wang}}, \bibinfo {author} {\bibfnamefont {D.~I.}\ \bibnamefont {Schuster}}, \bibinfo {author} {\bibfnamefont {S.~A.}\ \bibnamefont {Lyon}}, \bibinfo {author} {\bibfnamefont {J.}~\bibnamefont {Pollanen}},\ and\ \bibinfo {author} {\bibfnamefont {D.~G.}\ \bibnamefont {Rees}},\ }\bibfield  {title} {\bibinfo {title} {High-impedance resonators for strong coupling to an electron on helium},\ }\href@noop {} {\bibfield  {journal} {\bibinfo  {journal} {arXiv preprint arXiv:2410.19592}\ } (\bibinfo
  {year} {2024})}\BibitemShut {NoStop}%
\bibitem [{\citenamefont {Hazard}\ \emph {et~al.}(2019)\citenamefont {Hazard}, \citenamefont {Gyenis}, \citenamefont {Paolo}, \citenamefont {Asfaw}, \citenamefont {Lyon}, \citenamefont {Blais},\ and\ \citenamefont {Houck}}]{Hazard_nanowire_fluxonium}%
  \BibitemOpen
  \bibfield  {author} {\bibinfo {author} {\bibfnamefont {T.~M.}\ \bibnamefont {Hazard}}, \bibinfo {author} {\bibfnamefont {A.}~\bibnamefont {Gyenis}}, \bibinfo {author} {\bibfnamefont {A.~D.}\ \bibnamefont {Paolo}}, \bibinfo {author} {\bibfnamefont {A.~T.}\ \bibnamefont {Asfaw}}, \bibinfo {author} {\bibfnamefont {S.~A.}\ \bibnamefont {Lyon}}, \bibinfo {author} {\bibfnamefont {A.}~\bibnamefont {Blais}},\ and\ \bibinfo {author} {\bibfnamefont {A.~A.}\ \bibnamefont {Houck}},\ }\bibfield  {title} {\bibinfo {title} {Nanowire superinductance fluxonium qubit},\ }\href {https://doi.org/10.1103/PhysRevLett.122.010504} {\bibfield  {journal} {\bibinfo  {journal} {Phys. Rev. Lett.}\ }\textbf {\bibinfo {volume} {122}},\ \bibinfo {pages} {010504} (\bibinfo {year} {2019})}\BibitemShut {NoStop}%
\bibitem [{\citenamefont {Castellanos-Beltran}\ \emph {et~al.}(2008)\citenamefont {Castellanos-Beltran}, \citenamefont {Irwin}, \citenamefont {Hilton}, \citenamefont {Vale},\ and\ \citenamefont {Lehnert}}]{Castellanos-Beltran_metamaterial_amplifier}%
  \BibitemOpen
  \bibfield  {author} {\bibinfo {author} {\bibfnamefont {M.~A.}\ \bibnamefont {Castellanos-Beltran}}, \bibinfo {author} {\bibfnamefont {K.~D.}\ \bibnamefont {Irwin}}, \bibinfo {author} {\bibfnamefont {G.~C.}\ \bibnamefont {Hilton}}, \bibinfo {author} {\bibfnamefont {L.~R.}\ \bibnamefont {Vale}},\ and\ \bibinfo {author} {\bibfnamefont {K.~W.}\ \bibnamefont {Lehnert}},\ }\bibfield  {title} {\bibinfo {title} {Amplification and squeezing of quantum noise with a tunable {Josephson} metamaterial},\ }\href {https://doi.org/10.1038/nphys1090} {\bibfield  {journal} {\bibinfo  {journal} {Nature Phys}\ }\textbf {\bibinfo {volume} {4}},\ \bibinfo {pages} {929–931} (\bibinfo {year} {2008})}\BibitemShut {NoStop}%
\bibitem [{\citenamefont {Ranadive}\ \emph {et~al.}(2022)\citenamefont {Ranadive}, \citenamefont {Esposito}, \citenamefont {Planat}, \citenamefont {Bonet}, \citenamefont {Naud}, \citenamefont {Buisson}, \citenamefont {Guichard},\ and\ \citenamefont {Roch}}]{Ranadive_Kerr_reversal_amplifier}%
  \BibitemOpen
  \bibfield  {author} {\bibinfo {author} {\bibfnamefont {A.}~\bibnamefont {Ranadive}}, \bibinfo {author} {\bibfnamefont {M.}~\bibnamefont {Esposito}}, \bibinfo {author} {\bibfnamefont {L.}~\bibnamefont {Planat}}, \bibinfo {author} {\bibfnamefont {E.}~\bibnamefont {Bonet}}, \bibinfo {author} {\bibfnamefont {C.}~\bibnamefont {Naud}}, \bibinfo {author} {\bibfnamefont {O.}~\bibnamefont {Buisson}}, \bibinfo {author} {\bibfnamefont {W.}~\bibnamefont {Guichard}},\ and\ \bibinfo {author} {\bibfnamefont {N.}~\bibnamefont {Roch}},\ }\bibfield  {title} {\bibinfo {title} {Kerr reversal in {Josephson} meta-material and traveling wave parametric amplification},\ }\href {https://doi.org/10.1038/s41467-022-29375-5} {\bibfield  {journal} {\bibinfo  {journal} {Nat Commun}\ }\textbf {\bibinfo {volume} {13}},\ \bibinfo {pages} {1737} (\bibinfo {year} {2022})}\BibitemShut {NoStop}%
\bibitem [{\citenamefont {Peruzzo}\ \emph {et~al.}(2020)\citenamefont {Peruzzo}, \citenamefont {Trioni}, \citenamefont {Hassani}, \citenamefont {Zemlicka},\ and\ \citenamefont {Fink}}]{peruzzo2020surpassing}%
  \BibitemOpen
  \bibfield  {author} {\bibinfo {author} {\bibfnamefont {M.}~\bibnamefont {Peruzzo}}, \bibinfo {author} {\bibfnamefont {A.}~\bibnamefont {Trioni}}, \bibinfo {author} {\bibfnamefont {F.}~\bibnamefont {Hassani}}, \bibinfo {author} {\bibfnamefont {M.}~\bibnamefont {Zemlicka}},\ and\ \bibinfo {author} {\bibfnamefont {J.~M.}\ \bibnamefont {Fink}},\ }\bibfield  {title} {\bibinfo {title} {Surpassing the resistance quantum with a geometric superinductor},\ }\href {https://doi.org/10.1103/PhysRevApplied.14.044055} {\bibfield  {journal} {\bibinfo  {journal} {Phys. Rev. Appl.}\ }\textbf {\bibinfo {volume} {14}},\ \bibinfo {pages} {044055} (\bibinfo {year} {2020})}\BibitemShut {NoStop}%
\bibitem [{\citenamefont {Dunsworth}\ \emph {et~al.}(2017)\citenamefont {Dunsworth}, \citenamefont {Megrant}, \citenamefont {Quintana}, \citenamefont {Chen}, \citenamefont {Barends}, \citenamefont {Burkett}, \citenamefont {Foxen}, \citenamefont {Chen}, \citenamefont {Chiaro}, \citenamefont {Fowler}, \citenamefont {Graff}, \citenamefont {Jeffrey}, \citenamefont {Kelly}, \citenamefont {Lucero}, \citenamefont {Mutus}, \citenamefont {Neeley}, \citenamefont {Neill}, \citenamefont {Roushan}, \citenamefont {Sank}, \citenamefont {Vainsencher}, \citenamefont {Wenner}, \citenamefont {White},\ and\ \citenamefont {Martinis}}]{Dunsworth_bandaid_2017}%
  \BibitemOpen
  \bibfield  {author} {\bibinfo {author} {\bibfnamefont {A.}~\bibnamefont {Dunsworth}}, \bibinfo {author} {\bibfnamefont {A.}~\bibnamefont {Megrant}}, \bibinfo {author} {\bibfnamefont {C.}~\bibnamefont {Quintana}}, \bibinfo {author} {\bibfnamefont {Z.}~\bibnamefont {Chen}}, \bibinfo {author} {\bibfnamefont {R.}~\bibnamefont {Barends}}, \bibinfo {author} {\bibfnamefont {B.}~\bibnamefont {Burkett}}, \bibinfo {author} {\bibfnamefont {B.}~\bibnamefont {Foxen}}, \bibinfo {author} {\bibfnamefont {Y.}~\bibnamefont {Chen}}, \bibinfo {author} {\bibfnamefont {B.}~\bibnamefont {Chiaro}}, \bibinfo {author} {\bibfnamefont {A.}~\bibnamefont {Fowler}}, \bibinfo {author} {\bibfnamefont {R.}~\bibnamefont {Graff}}, \bibinfo {author} {\bibfnamefont {E.}~\bibnamefont {Jeffrey}}, \bibinfo {author} {\bibfnamefont {J.}~\bibnamefont {Kelly}}, \bibinfo {author} {\bibfnamefont {E.}~\bibnamefont {Lucero}}, \bibinfo {author} {\bibfnamefont {J.~Y.}\ \bibnamefont {Mutus}}, \bibinfo {author} {\bibfnamefont {M.}~\bibnamefont {Neeley}},
  \bibinfo {author} {\bibfnamefont {C.}~\bibnamefont {Neill}}, \bibinfo {author} {\bibfnamefont {P.}~\bibnamefont {Roushan}}, \bibinfo {author} {\bibfnamefont {D.}~\bibnamefont {Sank}}, \bibinfo {author} {\bibfnamefont {A.}~\bibnamefont {Vainsencher}}, \bibinfo {author} {\bibfnamefont {J.}~\bibnamefont {Wenner}}, \bibinfo {author} {\bibfnamefont {T.~C.}\ \bibnamefont {White}},\ and\ \bibinfo {author} {\bibfnamefont {J.~M.}\ \bibnamefont {Martinis}},\ }\bibfield  {title} {\bibinfo {title} {Characterization and reduction of capacitive loss induced by sub-micron josephson junction fabrication in superconducting qubits},\ }\href {https://doi.org/10.1063/1.4993577} {\bibfield  {journal} {\bibinfo  {journal} {Appl. Phys. Lett.}\ }\textbf {\bibinfo {volume} {111}},\ \bibinfo {pages} {022601} (\bibinfo {year} {2017})}\BibitemShut {NoStop}%
\bibitem [{\citenamefont {Somoroff}\ \emph {et~al.}(2023)\citenamefont {Somoroff}, \citenamefont {Ficheux}, \citenamefont {Mencia}, \citenamefont {Xiong}, \citenamefont {Kuzmin},\ and\ \citenamefont {Manucharyan}}]{Somoroff_ms_fluxonium}%
  \BibitemOpen
  \bibfield  {author} {\bibinfo {author} {\bibfnamefont {A.}~\bibnamefont {Somoroff}}, \bibinfo {author} {\bibfnamefont {Q.}~\bibnamefont {Ficheux}}, \bibinfo {author} {\bibfnamefont {R.~A.}\ \bibnamefont {Mencia}}, \bibinfo {author} {\bibfnamefont {H.}~\bibnamefont {Xiong}}, \bibinfo {author} {\bibfnamefont {R.}~\bibnamefont {Kuzmin}},\ and\ \bibinfo {author} {\bibfnamefont {V.~E.}\ \bibnamefont {Manucharyan}},\ }\bibfield  {title} {\bibinfo {title} {Millisecond coherence in a superconducting qubit},\ }\href {https://doi.org/10.1103/PhysRevLett.130.267001} {\bibfield  {journal} {\bibinfo  {journal} {Phys. Rev. Lett.}\ }\textbf {\bibinfo {volume} {130}},\ \bibinfo {pages} {267001} (\bibinfo {year} {2023})}\BibitemShut {NoStop}%
\bibitem [{\citenamefont {Murray}(2021)}]{Murray_mat_rev_2021}%
  \BibitemOpen
  \bibfield  {author} {\bibinfo {author} {\bibfnamefont {C.~E.}\ \bibnamefont {Murray}},\ }\bibfield  {title} {\bibinfo {title} {Material matters in superconducting qubits},\ }\href {https://doi.org/10.1016/j.mser.2021.100646} {\bibfield  {journal} {\bibinfo  {journal} {Mater. Sci. Eng. R Rep.}\ }\textbf {\bibinfo {volume} {146}},\ \bibinfo {pages} {100646} (\bibinfo {year} {2021})}\BibitemShut {NoStop}%
\bibitem [{\citenamefont {Chen}\ \emph {et~al.}(2024)\citenamefont {Chen}, \citenamefont {Owens}, \citenamefont {Putterman}, \citenamefont {Sch{\"a}fer},\ and\ \citenamefont {Painter}}]{chen2024phonon}%
  \BibitemOpen
  \bibfield  {author} {\bibinfo {author} {\bibfnamefont {M.}~\bibnamefont {Chen}}, \bibinfo {author} {\bibfnamefont {J.~C.}\ \bibnamefont {Owens}}, \bibinfo {author} {\bibfnamefont {H.}~\bibnamefont {Putterman}}, \bibinfo {author} {\bibfnamefont {M.}~\bibnamefont {Sch{\"a}fer}},\ and\ \bibinfo {author} {\bibfnamefont {O.}~\bibnamefont {Painter}},\ }\bibfield  {title} {\bibinfo {title} {Phonon engineering of atomic-scale defects in superconducting quantum circuits},\ }\href@noop {} {\bibfield  {journal} {\bibinfo  {journal} {Science Advances}\ }\textbf {\bibinfo {volume} {10}},\ \bibinfo {pages} {eado6240} (\bibinfo {year} {2024})}\BibitemShut {NoStop}%
\bibitem [{\citenamefont {Zhang}\ \emph {et~al.}(2024)\citenamefont {Zhang}, \citenamefont {Godeneli}, \citenamefont {He}, \citenamefont {Odeh}, \citenamefont {Zhou}, \citenamefont {Meesala},\ and\ \citenamefont {Sipahigil}}]{zhang2024acceptor}%
  \BibitemOpen
  \bibfield  {author} {\bibinfo {author} {\bibfnamefont {Z.-H.}\ \bibnamefont {Zhang}}, \bibinfo {author} {\bibfnamefont {K.}~\bibnamefont {Godeneli}}, \bibinfo {author} {\bibfnamefont {J.}~\bibnamefont {He}}, \bibinfo {author} {\bibfnamefont {M.}~\bibnamefont {Odeh}}, \bibinfo {author} {\bibfnamefont {H.}~\bibnamefont {Zhou}}, \bibinfo {author} {\bibfnamefont {S.}~\bibnamefont {Meesala}},\ and\ \bibinfo {author} {\bibfnamefont {A.}~\bibnamefont {Sipahigil}},\ }\bibfield  {title} {\bibinfo {title} {Acceptor-induced bulk dielectric loss in superconducting circuits on silicon},\ }\href {https://doi.org/10.1103/PhysRevX.14.041022} {\bibfield  {journal} {\bibinfo  {journal} {Phys. Rev. X}\ }\textbf {\bibinfo {volume} {14}},\ \bibinfo {pages} {041022} (\bibinfo {year} {2024})}\BibitemShut {NoStop}%
\bibitem [{\citenamefont {Catelani}\ \emph {et~al.}(2012)\citenamefont {Catelani}, \citenamefont {Nigg}, \citenamefont {Girvin}, \citenamefont {Schoelkopf},\ and\ \citenamefont {Glazman}}]{Catelani_decoherence_qp}%
  \BibitemOpen
  \bibfield  {author} {\bibinfo {author} {\bibfnamefont {G.}~\bibnamefont {Catelani}}, \bibinfo {author} {\bibfnamefont {S.~E.}\ \bibnamefont {Nigg}}, \bibinfo {author} {\bibfnamefont {S.~M.}\ \bibnamefont {Girvin}}, \bibinfo {author} {\bibfnamefont {R.~J.}\ \bibnamefont {Schoelkopf}},\ and\ \bibinfo {author} {\bibfnamefont {L.~I.}\ \bibnamefont {Glazman}},\ }\bibfield  {title} {\bibinfo {title} {Decoherence of superconducting qubits caused by quasiparticle tunneling},\ }\href {https://doi.org/10.1103/PhysRevB.86.184514} {\bibfield  {journal} {\bibinfo  {journal} {Phys. Rev. B}\ }\textbf {\bibinfo {volume} {86}},\ \bibinfo {pages} {184514} (\bibinfo {year} {2012})}\BibitemShut {NoStop}%
\bibitem [{\citenamefont {Glazman}\ and\ \citenamefont {Catelani}(2021)}]{Glazman_qp_lec_notes}%
  \BibitemOpen
  \bibfield  {author} {\bibinfo {author} {\bibfnamefont {L.~I.}\ \bibnamefont {Glazman}}\ and\ \bibinfo {author} {\bibfnamefont {G.}~\bibnamefont {Catelani}},\ }\bibfield  {title} {\bibinfo {title} {Bogoliubov quasiparticles in superconducting qubits},\ }\bibfield  {journal} {\bibinfo  {journal} {SciPost Phys. Lect. Notes}\ }\textbf {\bibinfo {volume} {31}},\ \href {https://doi.org/10.21468/SciPostPhysLectNotes.31} {10.21468/SciPostPhysLectNotes.31} (\bibinfo {year} {2021})\BibitemShut {NoStop}%
\bibitem [{\citenamefont {Wilen}\ \emph {et~al.}(2021)\citenamefont {Wilen}, \citenamefont {Abdullah}, \citenamefont {Kurinsky}, \citenamefont {Stanford}, \citenamefont {Cardani}, \citenamefont {d’Imperio}, \citenamefont {Tomei}, \citenamefont {Faoro}, \citenamefont {Ioffe}, \citenamefont {Liu}, \citenamefont {Opremcak}, \citenamefont {Christensen}, \citenamefont {DuBois},\ and\ \citenamefont {McDermott}}]{wilen2021correlated}%
  \BibitemOpen
  \bibfield  {author} {\bibinfo {author} {\bibfnamefont {C.~D.}\ \bibnamefont {Wilen}}, \bibinfo {author} {\bibfnamefont {S.}~\bibnamefont {Abdullah}}, \bibinfo {author} {\bibfnamefont {N.~A.}\ \bibnamefont {Kurinsky}}, \bibinfo {author} {\bibfnamefont {C.}~\bibnamefont {Stanford}}, \bibinfo {author} {\bibfnamefont {L.}~\bibnamefont {Cardani}}, \bibinfo {author} {\bibfnamefont {G.}~\bibnamefont {d’Imperio}}, \bibinfo {author} {\bibfnamefont {C.}~\bibnamefont {Tomei}}, \bibinfo {author} {\bibfnamefont {L.}~\bibnamefont {Faoro}}, \bibinfo {author} {\bibfnamefont {L.}~\bibnamefont {Ioffe}}, \bibinfo {author} {\bibfnamefont {C.~H.}\ \bibnamefont {Liu}}, \bibinfo {author} {\bibfnamefont {A.}~\bibnamefont {Opremcak}}, \bibinfo {author} {\bibfnamefont {B.~G.}\ \bibnamefont {Christensen}}, \bibinfo {author} {\bibfnamefont {J.}~\bibnamefont {DuBois}},\ and\ \bibinfo {author} {\bibfnamefont {R.}~\bibnamefont {McDermott}},\ }\bibfield  {title} {\bibinfo {title} {Correlated charge noise and relaxation errors in
  superconducting qubits},\ }\href {https://www.nature.com/articles/s41586-021-03557-5} {\bibfield  {journal} {\bibinfo  {journal} {Nature}\ }\textbf {\bibinfo {volume} {594}},\ \bibinfo {pages} {369–373} (\bibinfo {year} {2021})}\BibitemShut {NoStop}%
\bibitem [{\citenamefont {Iaia}\ \emph {et~al.}(2022)\citenamefont {Iaia}, \citenamefont {Ku}, \citenamefont {Ballard}, \citenamefont {Larson}, \citenamefont {Yelton}, \citenamefont {Liu}, \citenamefont {Patel}, \citenamefont {McDermott},\ and\ \citenamefont {Plourde}}]{iaia2022phonon}%
  \BibitemOpen
  \bibfield  {author} {\bibinfo {author} {\bibfnamefont {V.}~\bibnamefont {Iaia}}, \bibinfo {author} {\bibfnamefont {J.}~\bibnamefont {Ku}}, \bibinfo {author} {\bibfnamefont {A.}~\bibnamefont {Ballard}}, \bibinfo {author} {\bibfnamefont {C.}~\bibnamefont {Larson}}, \bibinfo {author} {\bibfnamefont {E.}~\bibnamefont {Yelton}}, \bibinfo {author} {\bibfnamefont {C.}~\bibnamefont {Liu}}, \bibinfo {author} {\bibfnamefont {S.}~\bibnamefont {Patel}}, \bibinfo {author} {\bibfnamefont {R.}~\bibnamefont {McDermott}},\ and\ \bibinfo {author} {\bibfnamefont {B.}~\bibnamefont {Plourde}},\ }\bibfield  {title} {\bibinfo {title} {Phonon downconversion to suppress correlated errors in superconducting qubits},\ }\href {https://www.nature.com/articles/s41467-022-33997-0} {\bibfield  {journal} {\bibinfo  {journal} {Nature Communications}\ }\textbf {\bibinfo {volume} {13}},\ \bibinfo {pages} {6425} (\bibinfo {year} {2022})}\BibitemShut {NoStop}%
\bibitem [{\citenamefont {Diamond}\ \emph {et~al.}(2022)\citenamefont {Diamond}, \citenamefont {Fatemi}, \citenamefont {Hays}, \citenamefont {Nho}, \citenamefont {Kurilovich}, \citenamefont {Connolly}, \citenamefont {Joshi}, \citenamefont {Serniak}, \citenamefont {Frunzio}, \citenamefont {Glazman},\ and\ \citenamefont {Devoret}}]{diamond2022distinguishing}%
  \BibitemOpen
  \bibfield  {author} {\bibinfo {author} {\bibfnamefont {S.}~\bibnamefont {Diamond}}, \bibinfo {author} {\bibfnamefont {V.}~\bibnamefont {Fatemi}}, \bibinfo {author} {\bibfnamefont {M.}~\bibnamefont {Hays}}, \bibinfo {author} {\bibfnamefont {H.}~\bibnamefont {Nho}}, \bibinfo {author} {\bibfnamefont {P.~D.}\ \bibnamefont {Kurilovich}}, \bibinfo {author} {\bibfnamefont {T.}~\bibnamefont {Connolly}}, \bibinfo {author} {\bibfnamefont {V.~R.}\ \bibnamefont {Joshi}}, \bibinfo {author} {\bibfnamefont {K.}~\bibnamefont {Serniak}}, \bibinfo {author} {\bibfnamefont {L.}~\bibnamefont {Frunzio}}, \bibinfo {author} {\bibfnamefont {L.~I.}\ \bibnamefont {Glazman}},\ and\ \bibinfo {author} {\bibfnamefont {M.~H.}\ \bibnamefont {Devoret}},\ }\bibfield  {title} {\bibinfo {title} {Distinguishing parity-switching mechanisms in a superconducting qubit},\ }\href {https://doi.org/10.1103/PRXQuantum.3.040304} {\bibfield  {journal} {\bibinfo  {journal} {PRX Quantum}\ }\textbf {\bibinfo {volume} {3}},\ \bibinfo {pages} {040304}
  (\bibinfo {year} {2022})}\BibitemShut {NoStop}%
\bibitem [{\citenamefont {Connolly}\ \emph {et~al.}(2024)\citenamefont {Connolly}, \citenamefont {Kurilovich}, \citenamefont {Diamond}, \citenamefont {Nho}, \citenamefont {B\o{}ttcher}, \citenamefont {Glazman}, \citenamefont {Fatemi},\ and\ \citenamefont {Devoret}}]{connolly2024coexistence}%
  \BibitemOpen
  \bibfield  {author} {\bibinfo {author} {\bibfnamefont {T.}~\bibnamefont {Connolly}}, \bibinfo {author} {\bibfnamefont {P.~D.}\ \bibnamefont {Kurilovich}}, \bibinfo {author} {\bibfnamefont {S.}~\bibnamefont {Diamond}}, \bibinfo {author} {\bibfnamefont {H.}~\bibnamefont {Nho}}, \bibinfo {author} {\bibfnamefont {C.~G.~L.}\ \bibnamefont {B\o{}ttcher}}, \bibinfo {author} {\bibfnamefont {L.~I.}\ \bibnamefont {Glazman}}, \bibinfo {author} {\bibfnamefont {V.}~\bibnamefont {Fatemi}},\ and\ \bibinfo {author} {\bibfnamefont {M.~H.}\ \bibnamefont {Devoret}},\ }\bibfield  {title} {\bibinfo {title} {Coexistence of nonequilibrium density and equilibrium energy distribution of quasiparticles in a superconducting qubit},\ }\href {https://doi.org/10.1103/PhysRevLett.132.217001} {\bibfield  {journal} {\bibinfo  {journal} {Phys. Rev. Lett.}\ }\textbf {\bibinfo {volume} {132}},\ \bibinfo {pages} {217001} (\bibinfo {year} {2024})}\BibitemShut {NoStop}%
\bibitem [{\citenamefont {Hashim}\ \emph {et~al.}(2024)\citenamefont {Hashim}, \citenamefont {Yuan}, \citenamefont {Gokhale}, \citenamefont {Chen}, \citenamefont {Juenger}, \citenamefont {Fruitwala}, \citenamefont {Xu}, \citenamefont {Huang}, \citenamefont {Jiang},\ and\ \citenamefont {Siddiqi}}]{Hashim2024}%
  \BibitemOpen
  \bibfield  {author} {\bibinfo {author} {\bibfnamefont {A.}~\bibnamefont {Hashim}}, \bibinfo {author} {\bibfnamefont {M.}~\bibnamefont {Yuan}}, \bibinfo {author} {\bibfnamefont {P.}~\bibnamefont {Gokhale}}, \bibinfo {author} {\bibfnamefont {L.}~\bibnamefont {Chen}}, \bibinfo {author} {\bibfnamefont {C.}~\bibnamefont {Juenger}}, \bibinfo {author} {\bibfnamefont {N.}~\bibnamefont {Fruitwala}}, \bibinfo {author} {\bibfnamefont {Y.}~\bibnamefont {Xu}}, \bibinfo {author} {\bibfnamefont {G.}~\bibnamefont {Huang}}, \bibinfo {author} {\bibfnamefont {L.}~\bibnamefont {Jiang}},\ and\ \bibinfo {author} {\bibfnamefont {I.}~\bibnamefont {Siddiqi}},\ }\bibfield  {title} {\bibinfo {title} {Efficient generation of multi-partite entanglement between non-local superconducting qubits using classical feedback},\ }\bibfield  {journal} {\bibinfo  {journal} {arXiv}\ }\href {https://doi.org/10.48550/ARXIV.2403.18768} {10.48550/ARXIV.2403.18768} (\bibinfo {year} {2024}),\ \Eprint {https://arxiv.org/abs/2403.18768} {2403.18768
  [quant-ph]} \BibitemShut {NoStop}%
\bibitem [{\citenamefont {Altoé}\ \emph {et~al.}(2022)\citenamefont {Altoé}, \citenamefont {Banerjee}, \citenamefont {Berk}, \citenamefont {Hajr}, \citenamefont {Schwartzberg}, \citenamefont {Song}, \citenamefont {Alghadeer}, \citenamefont {Aloni}, \citenamefont {Elowson}, \citenamefont {Kreikebaum}, \citenamefont {Wong}, \citenamefont {Griffin}, \citenamefont {Rao}, \citenamefont {Weber-Bargioni}, \citenamefont {Minor}, \citenamefont {Santiago}, \citenamefont {Cabrini}, \citenamefont {Siddiqi},\ and\ \citenamefont {Ogletree}}]{Altoe2022}%
  \BibitemOpen
  \bibfield  {author} {\bibinfo {author} {\bibfnamefont {M.~V.~P.}\ \bibnamefont {Altoé}}, \bibinfo {author} {\bibfnamefont {A.}~\bibnamefont {Banerjee}}, \bibinfo {author} {\bibfnamefont {C.}~\bibnamefont {Berk}}, \bibinfo {author} {\bibfnamefont {A.}~\bibnamefont {Hajr}}, \bibinfo {author} {\bibfnamefont {A.}~\bibnamefont {Schwartzberg}}, \bibinfo {author} {\bibfnamefont {C.}~\bibnamefont {Song}}, \bibinfo {author} {\bibfnamefont {M.}~\bibnamefont {Alghadeer}}, \bibinfo {author} {\bibfnamefont {S.}~\bibnamefont {Aloni}}, \bibinfo {author} {\bibfnamefont {M.~J.}\ \bibnamefont {Elowson}}, \bibinfo {author} {\bibfnamefont {J.~M.}\ \bibnamefont {Kreikebaum}}, \bibinfo {author} {\bibfnamefont {E.~K.}\ \bibnamefont {Wong}}, \bibinfo {author} {\bibfnamefont {S.~M.}\ \bibnamefont {Griffin}}, \bibinfo {author} {\bibfnamefont {S.}~\bibnamefont {Rao}}, \bibinfo {author} {\bibfnamefont {A.}~\bibnamefont {Weber-Bargioni}}, \bibinfo {author} {\bibfnamefont {A.~M.}\ \bibnamefont {Minor}}, \bibinfo {author} {\bibfnamefont
  {D.~I.}\ \bibnamefont {Santiago}}, \bibinfo {author} {\bibfnamefont {S.}~\bibnamefont {Cabrini}}, \bibinfo {author} {\bibfnamefont {I.}~\bibnamefont {Siddiqi}},\ and\ \bibinfo {author} {\bibfnamefont {D.~F.}\ \bibnamefont {Ogletree}},\ }\bibfield  {title} {\bibinfo {title} {Localization and mitigation of loss in niobium superconducting circuits},\ }\href {https://doi.org/10.1103/prxquantum.3.020312} {\bibfield  {journal} {\bibinfo  {journal} {PRX Quantum}\ }\textbf {\bibinfo {volume} {3}},\ \bibinfo {pages} {020312} (\bibinfo {year} {2022})}\BibitemShut {NoStop}%
\bibitem [{\citenamefont {Dolan}(1977)}]{Dolan1977}%
  \BibitemOpen
  \bibfield  {author} {\bibinfo {author} {\bibfnamefont {G.~J.}\ \bibnamefont {Dolan}},\ }\bibfield  {title} {\bibinfo {title} {Offset masks for lift-off photoprocessing},\ }\href {https://doi.org/10.1063/1.89690} {\bibfield  {journal} {\bibinfo  {journal} {Applied Physics Letters}\ }\textbf {\bibinfo {volume} {31}},\ \bibinfo {pages} {337} (\bibinfo {year} {1977})}\BibitemShut {NoStop}%
\bibitem [{\citenamefont {Winters}\ and\ \citenamefont {Coburn}(1979)}]{Winters_XeF2}%
  \BibitemOpen
  \bibfield  {author} {\bibinfo {author} {\bibfnamefont {H.~F.}\ \bibnamefont {Winters}}\ and\ \bibinfo {author} {\bibfnamefont {J.~W.}\ \bibnamefont {Coburn}},\ }\bibfield  {title} {\bibinfo {title} {The etching of silicon with {$\text{XeF}_2$} vapor},\ }\href {https://doi.org/10.1063/1.90562} {\bibfield  {journal} {\bibinfo  {journal} {Appl. Phys. Lett.}\ }\textbf {\bibinfo {volume} {34}},\ \bibinfo {pages} {70–73} (\bibinfo {year} {1979})}\BibitemShut {NoStop}%
\bibitem [{\citenamefont {Chang}\ \emph {et~al.}(1995)\citenamefont {Chang}, \citenamefont {Yeh}, \citenamefont {Lin}, \citenamefont {Chu}, \citenamefont {Hoffman}, \citenamefont {Kruglick}, \citenamefont {Pister},\ and\ \citenamefont {Hecht}}]{Chang_XeF2}%
  \BibitemOpen
  \bibfield  {author} {\bibinfo {author} {\bibfnamefont {F.~I.}\ \bibnamefont {Chang}}, \bibinfo {author} {\bibfnamefont {R.}~\bibnamefont {Yeh}}, \bibinfo {author} {\bibfnamefont {G.}~\bibnamefont {Lin}}, \bibinfo {author} {\bibfnamefont {P.~B.}\ \bibnamefont {Chu}}, \bibinfo {author} {\bibfnamefont {E.}~\bibnamefont {Hoffman}}, \bibinfo {author} {\bibfnamefont {E.~J.~J.}\ \bibnamefont {Kruglick}}, \bibinfo {author} {\bibfnamefont {K.~S.~J.}\ \bibnamefont {Pister}},\ and\ \bibinfo {author} {\bibfnamefont {M.~H.}\ \bibnamefont {Hecht}},\ }\bibfield  {title} {\bibinfo {title} {Gas-phase silicon micromachining with xenon difluoride},\ }\bibfield  {journal} {\bibinfo  {journal} {Proc. SPIE 2641, Microelectronic Structures and Microelectromechanical Devices for Optical Processing and Multimedia Applications}\ }\href {https://doi.org/10.1117/12.220933} {10.1117/12.220933} (\bibinfo {year} {1995})\BibitemShut {NoStop}%
\bibitem [{\citenamefont {Chan}\ \emph {et~al.}(1999)\citenamefont {Chan}, \citenamefont {Brown}, \citenamefont {Lawson}, \citenamefont {Robinson}, \citenamefont {Ma},\ and\ \citenamefont {Strembicke}}]{Chan_XeF2}%
  \BibitemOpen
  \bibfield  {author} {\bibinfo {author} {\bibfnamefont {I.~W.~T.}\ \bibnamefont {Chan}}, \bibinfo {author} {\bibfnamefont {K.~B.}\ \bibnamefont {Brown}}, \bibinfo {author} {\bibfnamefont {R.~P.~W.}\ \bibnamefont {Lawson}}, \bibinfo {author} {\bibfnamefont {A.~M.}\ \bibnamefont {Robinson}}, \bibinfo {author} {\bibfnamefont {Y.}~\bibnamefont {Ma}},\ and\ \bibinfo {author} {\bibfnamefont {D.}~\bibnamefont {Strembicke}},\ }\bibfield  {title} {\bibinfo {title} {Gas phase pulse etching of silicon for {MEMS} with xenon difluoride},\ }\href {https://doi.org/10.1109/CCECE.1999.804962} {\bibfield  {journal} {\bibinfo  {journal} {Engineering Solutions for the Next Millenium. 1999 IEEE Canadian Conference on Electrical and Computer Engineering (Cat. No.99TH8411)}\ }\textbf {\bibinfo {volume} {1633}},\ \bibinfo {pages} {1637–1642} (\bibinfo {year} {1999})}\BibitemShut {NoStop}%
\bibitem [{\citenamefont {McRae}\ \emph {et~al.}(2020)\citenamefont {McRae}, \citenamefont {Wang}, \citenamefont {Gao}, \citenamefont {Vissers}, \citenamefont {Brecht}, \citenamefont {Dunsworth}, \citenamefont {Pappas},\ and\ \citenamefont {Mutus}}]{McRae_res}%
  \BibitemOpen
  \bibfield  {author} {\bibinfo {author} {\bibfnamefont {C.~R.~H.}\ \bibnamefont {McRae}}, \bibinfo {author} {\bibfnamefont {H.}~\bibnamefont {Wang}}, \bibinfo {author} {\bibfnamefont {J.}~\bibnamefont {Gao}}, \bibinfo {author} {\bibfnamefont {M.~R.}\ \bibnamefont {Vissers}}, \bibinfo {author} {\bibfnamefont {T.}~\bibnamefont {Brecht}}, \bibinfo {author} {\bibfnamefont {A.}~\bibnamefont {Dunsworth}}, \bibinfo {author} {\bibfnamefont {D.~P.}\ \bibnamefont {Pappas}},\ and\ \bibinfo {author} {\bibfnamefont {J.}~\bibnamefont {Mutus}},\ }\bibfield  {title} {\bibinfo {title} {Materials loss measurements using superconducting microwave resonators},\ }\href {https://doi.org/10.1063/5.0017378} {\bibfield  {journal} {\bibinfo  {journal} {Rev. Sci. Instrum.}\ }\textbf {\bibinfo {volume} {91}},\ \bibinfo {pages} {091101} (\bibinfo {year} {2020})}\BibitemShut {NoStop}%
\bibitem [{\citenamefont {Masluk}\ \emph {et~al.}(2012)\citenamefont {Masluk}, \citenamefont {Pop}, \citenamefont {Kamal}, \citenamefont {Minev},\ and\ \citenamefont {Devoret}}]{Masluk_JJ_array}%
  \BibitemOpen
  \bibfield  {author} {\bibinfo {author} {\bibfnamefont {N.~A.}\ \bibnamefont {Masluk}}, \bibinfo {author} {\bibfnamefont {I.~M.}\ \bibnamefont {Pop}}, \bibinfo {author} {\bibfnamefont {A.}~\bibnamefont {Kamal}}, \bibinfo {author} {\bibfnamefont {Z.~K.}\ \bibnamefont {Minev}},\ and\ \bibinfo {author} {\bibfnamefont {M.~H.}\ \bibnamefont {Devoret}},\ }\bibfield  {title} {\bibinfo {title} {Microwave characterization of {J}osephson junction arrays: Implementing a low loss superinductance},\ }\href {https://doi.org/10.1103/PhysRevLett.109.137002} {\bibfield  {journal} {\bibinfo  {journal} {Phys. Rev. Lett.}\ }\textbf {\bibinfo {volume} {109}},\ \bibinfo {pages} {137002} (\bibinfo {year} {2012})}\BibitemShut {NoStop}%
\bibitem [{\citenamefont {Weißl}\ \emph {et~al.}(2015)\citenamefont {Weißl}, \citenamefont {Küng}, \citenamefont {Dumur}, \citenamefont {Feofanov}, \citenamefont {Matei}, \citenamefont {Naud}, \citenamefont {Buisson}, \citenamefont {Hekking},\ and\ \citenamefont {Guichard}}]{Weissl_Kerr_coefficients}%
  \BibitemOpen
  \bibfield  {author} {\bibinfo {author} {\bibfnamefont {T.}~\bibnamefont {Weißl}}, \bibinfo {author} {\bibfnamefont {B.}~\bibnamefont {Küng}}, \bibinfo {author} {\bibfnamefont {E.}~\bibnamefont {Dumur}}, \bibinfo {author} {\bibfnamefont {A.~K.}\ \bibnamefont {Feofanov}}, \bibinfo {author} {\bibfnamefont {I.}~\bibnamefont {Matei}}, \bibinfo {author} {\bibfnamefont {C.}~\bibnamefont {Naud}}, \bibinfo {author} {\bibfnamefont {O.}~\bibnamefont {Buisson}}, \bibinfo {author} {\bibfnamefont {F.~W.~J.}\ \bibnamefont {Hekking}},\ and\ \bibinfo {author} {\bibfnamefont {W.}~\bibnamefont {Guichard}},\ }\bibfield  {title} {\bibinfo {title} {Kerr coefficients of plasma resonances in {Josephson} junction chains},\ }\href {https://doi.org/10.1103/PhysRevB.92.104508} {\bibfield  {journal} {\bibinfo  {journal} {Phys. Rev. B}\ }\textbf {\bibinfo {volume} {92}},\ \bibinfo {pages} {104508} (\bibinfo {year} {2015})}\BibitemShut {NoStop}%
\bibitem [{\citenamefont {Chu}\ \emph {et~al.}(2016)\citenamefont {Chu}, \citenamefont {Axline}, \citenamefont {Wang}, \citenamefont {Brecht}, \citenamefont {Gao}, \citenamefont {Frunzio},\ and\ \citenamefont {Schoelkopf}}]{Chu_micromachining}%
  \BibitemOpen
  \bibfield  {author} {\bibinfo {author} {\bibfnamefont {Y.}~\bibnamefont {Chu}}, \bibinfo {author} {\bibfnamefont {C.}~\bibnamefont {Axline}}, \bibinfo {author} {\bibfnamefont {C.}~\bibnamefont {Wang}}, \bibinfo {author} {\bibfnamefont {T.}~\bibnamefont {Brecht}}, \bibinfo {author} {\bibfnamefont {Y.~Y.}\ \bibnamefont {Gao}}, \bibinfo {author} {\bibfnamefont {L.}~\bibnamefont {Frunzio}},\ and\ \bibinfo {author} {\bibfnamefont {R.~J.}\ \bibnamefont {Schoelkopf}},\ }\bibfield  {title} {\bibinfo {title} {Suspending superconducting qubits by silicon micromachining},\ }\href {https://doi.org/10.1063/1.4962327} {\bibfield  {journal} {\bibinfo  {journal} {Appl. Phys. Lett.}\ }\textbf {\bibinfo {volume} {109}},\ \bibinfo {pages} {112601} (\bibinfo {year} {2016})}\BibitemShut {NoStop}%
\bibitem [{\citenamefont {Wenner}\ \emph {et~al.}(2011)\citenamefont {Wenner}, \citenamefont {Barends}, \citenamefont {Bialczak}, \citenamefont {Chen}, \citenamefont {Kelly}, \citenamefont {Lucero}, \citenamefont {Mariantoni}, \citenamefont {Megrant}, \citenamefont {O'Malley}, \citenamefont {Sank}, \citenamefont {Vainsencher}, \citenamefont {Wang}, \citenamefont {White}, \citenamefont {Yin}, \citenamefont {Zhao}, \citenamefont {Cleland},\ and\ \citenamefont {Martinis}}]{Wenner_res_sim}%
  \BibitemOpen
  \bibfield  {author} {\bibinfo {author} {\bibfnamefont {J.}~\bibnamefont {Wenner}}, \bibinfo {author} {\bibfnamefont {R.}~\bibnamefont {Barends}}, \bibinfo {author} {\bibfnamefont {R.~C.}\ \bibnamefont {Bialczak}}, \bibinfo {author} {\bibfnamefont {Y.}~\bibnamefont {Chen}}, \bibinfo {author} {\bibfnamefont {J.}~\bibnamefont {Kelly}}, \bibinfo {author} {\bibfnamefont {E.}~\bibnamefont {Lucero}}, \bibinfo {author} {\bibfnamefont {M.}~\bibnamefont {Mariantoni}}, \bibinfo {author} {\bibfnamefont {A.}~\bibnamefont {Megrant}}, \bibinfo {author} {\bibfnamefont {P.~J.~J.}\ \bibnamefont {O'Malley}}, \bibinfo {author} {\bibfnamefont {D.}~\bibnamefont {Sank}}, \bibinfo {author} {\bibfnamefont {A.}~\bibnamefont {Vainsencher}}, \bibinfo {author} {\bibfnamefont {H.}~\bibnamefont {Wang}}, \bibinfo {author} {\bibfnamefont {T.~C.}\ \bibnamefont {White}}, \bibinfo {author} {\bibfnamefont {Y.}~\bibnamefont {Yin}}, \bibinfo {author} {\bibfnamefont {J.}~\bibnamefont {Zhao}}, \bibinfo {author} {\bibfnamefont {A.~N.}\ \bibnamefont
  {Cleland}},\ and\ \bibinfo {author} {\bibfnamefont {J.~M.}\ \bibnamefont {Martinis}},\ }\bibfield  {title} {\bibinfo {title} {Surface loss simulations of superconducting coplanar waveguide resonators},\ }\href {https://doi.org/10.1063/1.3637047} {\bibfield  {journal} {\bibinfo  {journal} {Appl. Phys. Lett.}\ }\textbf {\bibinfo {volume} {99}},\ \bibinfo {pages} {113513} (\bibinfo {year} {2011})}\BibitemShut {NoStop}%
\bibitem [{\citenamefont {Woods}\ \emph {et~al.}(2019)\citenamefont {Woods}, \citenamefont {Calusine}, \citenamefont {Melville}, \citenamefont {Sevi}, \citenamefont {Golden}, \citenamefont {Kim}, \citenamefont {Rosenberg}, \citenamefont {Yoder},\ and\ \citenamefont {Oliver}}]{Woods_res}%
  \BibitemOpen
  \bibfield  {author} {\bibinfo {author} {\bibfnamefont {W.}~\bibnamefont {Woods}}, \bibinfo {author} {\bibfnamefont {G.}~\bibnamefont {Calusine}}, \bibinfo {author} {\bibfnamefont {A.}~\bibnamefont {Melville}}, \bibinfo {author} {\bibfnamefont {A.}~\bibnamefont {Sevi}}, \bibinfo {author} {\bibfnamefont {E.}~\bibnamefont {Golden}}, \bibinfo {author} {\bibfnamefont {D.~K.}\ \bibnamefont {Kim}}, \bibinfo {author} {\bibfnamefont {D.}~\bibnamefont {Rosenberg}}, \bibinfo {author} {\bibfnamefont {J.~L.}\ \bibnamefont {Yoder}},\ and\ \bibinfo {author} {\bibfnamefont {W.~D.}\ \bibnamefont {Oliver}},\ }\bibfield  {title} {\bibinfo {title} {Determining interface dielectric losses in superconducting coplanar-waveguide resonators},\ }\href {https://doi.org/10.1103/PhysRevApplied.12.014012} {\bibfield  {journal} {\bibinfo  {journal} {Phys. Rev. Applied}\ }\textbf {\bibinfo {volume} {12}},\ \bibinfo {pages} {014012} (\bibinfo {year} {2019})}\BibitemShut {NoStop}%
\bibitem [{\citenamefont {Joshi}\ \emph {et~al.}(2022)\citenamefont {Joshi}, \citenamefont {Chen}, \citenamefont {LeDuc}, \citenamefont {Day},\ and\ \citenamefont {Mirhosseini}}]{Joshi_TiN_Kerr}%
  \BibitemOpen
  \bibfield  {author} {\bibinfo {author} {\bibfnamefont {C.}~\bibnamefont {Joshi}}, \bibinfo {author} {\bibfnamefont {W.}~\bibnamefont {Chen}}, \bibinfo {author} {\bibfnamefont {H.~G.}\ \bibnamefont {LeDuc}}, \bibinfo {author} {\bibfnamefont {P.~K.}\ \bibnamefont {Day}},\ and\ \bibinfo {author} {\bibfnamefont {M.}~\bibnamefont {Mirhosseini}},\ }\bibfield  {title} {\bibinfo {title} {Strong kinetic-inductance kerr nonlinearity with titanium nitride nanowires},\ }\href {https://doi.org/10.1103/PhysRevApplied.18.064088} {\bibfield  {journal} {\bibinfo  {journal} {Phys. Rev. Applied}\ }\textbf {\bibinfo {volume} {18}},\ \bibinfo {pages} {064088} (\bibinfo {year} {2022})}\BibitemShut {NoStop}%
\bibitem [{\citenamefont {Krupko}\ \emph {et~al.}(2018)\citenamefont {Krupko}, \citenamefont {Nguyen}, \citenamefont {Weißl}, \citenamefont {Dumur}, \citenamefont {Puertas}, \citenamefont {Dassonneville}, \citenamefont {Naud}, \citenamefont {Hekking}, \citenamefont {Basko}, \citenamefont {Buisson}, \citenamefont {Roch},\ and\ \citenamefont {Hasch-Guichard}}]{Krupko_Kerr_metamaterial_2018}%
  \BibitemOpen
  \bibfield  {author} {\bibinfo {author} {\bibfnamefont {Y.}~\bibnamefont {Krupko}}, \bibinfo {author} {\bibfnamefont {V.~D.}\ \bibnamefont {Nguyen}}, \bibinfo {author} {\bibfnamefont {T.}~\bibnamefont {Weißl}}, \bibinfo {author} {\bibfnamefont {E.}~\bibnamefont {Dumur}}, \bibinfo {author} {\bibfnamefont {J.}~\bibnamefont {Puertas}}, \bibinfo {author} {\bibfnamefont {R.}~\bibnamefont {Dassonneville}}, \bibinfo {author} {\bibfnamefont {C.}~\bibnamefont {Naud}}, \bibinfo {author} {\bibfnamefont {F.~W.~J.}\ \bibnamefont {Hekking}}, \bibinfo {author} {\bibfnamefont {D.~M.}\ \bibnamefont {Basko}}, \bibinfo {author} {\bibfnamefont {O.}~\bibnamefont {Buisson}}, \bibinfo {author} {\bibfnamefont {N.}~\bibnamefont {Roch}},\ and\ \bibinfo {author} {\bibfnamefont {W.}~\bibnamefont {Hasch-Guichard}},\ }\bibfield  {title} {\bibinfo {title} {Kerr nonlinearity in a superconducting josephson metamaterial},\ }\href {https://doi.org/10.1103/PhysRevB.98.094516} {\bibfield  {journal} {\bibinfo  {journal} {Phys. Rev. B}\ }\textbf
  {\bibinfo {volume} {98}},\ \bibinfo {pages} {094516} (\bibinfo {year} {2018})}\BibitemShut {NoStop}%
\bibitem [{\citenamefont {Fazio}\ and\ \citenamefont {van~der Zant}(2001)}]{Fazio_SC_networks}%
  \BibitemOpen
  \bibfield  {author} {\bibinfo {author} {\bibfnamefont {R.}~\bibnamefont {Fazio}}\ and\ \bibinfo {author} {\bibfnamefont {H.}~\bibnamefont {van~der Zant}},\ }\bibfield  {title} {\bibinfo {title} {Quantum phase transitions and vortex dynamics in superconducting networks},\ }\href {https://doi.org/10.1016/S0370-1573(01)00022-9} {\bibfield  {journal} {\bibinfo  {journal} {Physics Reports}\ }\textbf {\bibinfo {volume} {355}},\ \bibinfo {pages} {4} (\bibinfo {year} {2001})}\BibitemShut {NoStop}%
\bibitem [{\citenamefont {Vool}\ and\ \citenamefont {Devoret}(2017)}]{Vool_rev}%
  \BibitemOpen
  \bibfield  {author} {\bibinfo {author} {\bibfnamefont {U.}~\bibnamefont {Vool}}\ and\ \bibinfo {author} {\bibfnamefont {M.}~\bibnamefont {Devoret}},\ }\bibfield  {title} {\bibinfo {title} {Introduction to quantum electromagnetic circuits},\ }\href {https://doi.org/10.1002/cta.2359} {\bibfield  {journal} {\bibinfo  {journal} {International Journal of Circuit Theory and Applications}\ }\textbf {\bibinfo {volume} {45}},\ \bibinfo {pages} {7} (\bibinfo {year} {2017})}\BibitemShut {NoStop}%
\bibitem [{\citenamefont {Ding}\ \emph {et~al.}(2023)\citenamefont {Ding}, \citenamefont {Hays}, \citenamefont {Sung}, \citenamefont {Kannan}, \citenamefont {An}, \citenamefont {Di~Paolo}, \citenamefont {Karamlou}, \citenamefont {Hazard}, \citenamefont {Azar}, \citenamefont {Kim}, \citenamefont {Niedzielski}, \citenamefont {Melville}, \citenamefont {Schwartz}, \citenamefont {Yoder}, \citenamefont {Orlando}, \citenamefont {Gustavsson}, \citenamefont {Grover}, \citenamefont {Serniak},\ and\ \citenamefont {Oliver}}]{Ding2023}%
  \BibitemOpen
  \bibfield  {author} {\bibinfo {author} {\bibfnamefont {L.}~\bibnamefont {Ding}}, \bibinfo {author} {\bibfnamefont {M.}~\bibnamefont {Hays}}, \bibinfo {author} {\bibfnamefont {Y.}~\bibnamefont {Sung}}, \bibinfo {author} {\bibfnamefont {B.}~\bibnamefont {Kannan}}, \bibinfo {author} {\bibfnamefont {J.}~\bibnamefont {An}}, \bibinfo {author} {\bibfnamefont {A.}~\bibnamefont {Di~Paolo}}, \bibinfo {author} {\bibfnamefont {A.~H.}\ \bibnamefont {Karamlou}}, \bibinfo {author} {\bibfnamefont {T.~M.}\ \bibnamefont {Hazard}}, \bibinfo {author} {\bibfnamefont {K.}~\bibnamefont {Azar}}, \bibinfo {author} {\bibfnamefont {D.~K.}\ \bibnamefont {Kim}}, \bibinfo {author} {\bibfnamefont {B.~M.}\ \bibnamefont {Niedzielski}}, \bibinfo {author} {\bibfnamefont {A.}~\bibnamefont {Melville}}, \bibinfo {author} {\bibfnamefont {M.~E.}\ \bibnamefont {Schwartz}}, \bibinfo {author} {\bibfnamefont {J.~L.}\ \bibnamefont {Yoder}}, \bibinfo {author} {\bibfnamefont {T.~P.}\ \bibnamefont {Orlando}}, \bibinfo {author} {\bibfnamefont
  {S.}~\bibnamefont {Gustavsson}}, \bibinfo {author} {\bibfnamefont {J.~A.}\ \bibnamefont {Grover}}, \bibinfo {author} {\bibfnamefont {K.}~\bibnamefont {Serniak}},\ and\ \bibinfo {author} {\bibfnamefont {W.~D.}\ \bibnamefont {Oliver}},\ }\bibfield  {title} {\bibinfo {title} {High-fidelity, frequency-flexible two-qubit fluxonium gates with a transmon coupler},\ }\href {https://doi.org/10.1103/physrevx.13.031035} {\bibfield  {journal} {\bibinfo  {journal} {Physical Review X}\ }\textbf {\bibinfo {volume} {13}},\ \bibinfo {pages} {031035} (\bibinfo {year} {2023})}\BibitemShut {NoStop}%
\bibitem [{\citenamefont {Kreikebaum}\ \emph {et~al.}(2020)\citenamefont {Kreikebaum}, \citenamefont {O’Brien}, \citenamefont {Morvan},\ and\ \citenamefont {Siddiqi}}]{Kreikebaum2020}%
  \BibitemOpen
  \bibfield  {author} {\bibinfo {author} {\bibfnamefont {J.~M.}\ \bibnamefont {Kreikebaum}}, \bibinfo {author} {\bibfnamefont {K.~P.}\ \bibnamefont {O’Brien}}, \bibinfo {author} {\bibfnamefont {A.}~\bibnamefont {Morvan}},\ and\ \bibinfo {author} {\bibfnamefont {I.}~\bibnamefont {Siddiqi}},\ }\bibfield  {title} {\bibinfo {title} {Improving wafer-scale {Josephson} junction resistance variation in superconducting quantum coherent circuits},\ }\href {https://doi.org/10.1088/1361-6668/ab8617} {\bibfield  {journal} {\bibinfo  {journal} {Superconductor Science and Technology}\ }\textbf {\bibinfo {volume} {33}},\ \bibinfo {pages} {06LT02} (\bibinfo {year} {2020})}\BibitemShut {NoStop}%
\bibitem [{\citenamefont {Mencia}\ \emph {et~al.}(2024)\citenamefont {Mencia}, \citenamefont {Lin}, \citenamefont {Cho}, \citenamefont {Vavilov},\ and\ \citenamefont {Manucharyan}}]{Mencia_integer_fluxonium}%
  \BibitemOpen
  \bibfield  {author} {\bibinfo {author} {\bibfnamefont {R.~A.}\ \bibnamefont {Mencia}}, \bibinfo {author} {\bibfnamefont {W.-J.}\ \bibnamefont {Lin}}, \bibinfo {author} {\bibfnamefont {H.}~\bibnamefont {Cho}}, \bibinfo {author} {\bibfnamefont {M.~G.}\ \bibnamefont {Vavilov}},\ and\ \bibinfo {author} {\bibfnamefont {V.~E.}\ \bibnamefont {Manucharyan}},\ }\bibfield  {title} {\bibinfo {title} {Integer fluxonium qubit},\ }\href@noop {} {\bibfield  {journal} {\bibinfo  {journal} {arXiv preprint arXiv:2403.16780}\ } (\bibinfo {year} {2024})}\BibitemShut {NoStop}%
\bibitem [{\citenamefont {Hertzberg}\ \emph {et~al.}(2021)\citenamefont {Hertzberg}, \citenamefont {Zhang}, \citenamefont {Rosenblatt}, \citenamefont {Magesan}, \citenamefont {Smolin}, \citenamefont {Yau}, \citenamefont {Adiga}, \citenamefont {Sandberg}, \citenamefont {Brink}, \citenamefont {Chow},\ and\ \citenamefont {Orcutt}}]{Hertzberg_laser_annealing}%
  \BibitemOpen
  \bibfield  {author} {\bibinfo {author} {\bibfnamefont {J.~B.}\ \bibnamefont {Hertzberg}}, \bibinfo {author} {\bibfnamefont {E.~J.}\ \bibnamefont {Zhang}}, \bibinfo {author} {\bibfnamefont {S.}~\bibnamefont {Rosenblatt}}, \bibinfo {author} {\bibfnamefont {E.}~\bibnamefont {Magesan}}, \bibinfo {author} {\bibfnamefont {J.~A.}\ \bibnamefont {Smolin}}, \bibinfo {author} {\bibfnamefont {J.-B.}\ \bibnamefont {Yau}}, \bibinfo {author} {\bibfnamefont {V.~P.}\ \bibnamefont {Adiga}}, \bibinfo {author} {\bibfnamefont {M.}~\bibnamefont {Sandberg}}, \bibinfo {author} {\bibfnamefont {M.}~\bibnamefont {Brink}}, \bibinfo {author} {\bibfnamefont {J.~M.}\ \bibnamefont {Chow}},\ and\ \bibinfo {author} {\bibfnamefont {J.~S.}\ \bibnamefont {Orcutt}},\ }\bibfield  {title} {\bibinfo {title} {Laser-annealing {Josephson} junctions for yielding scaled-up superconducting quantum processors},\ }\href {https://doi.org/10.1038/s41534-021-00464-5} {\bibfield  {journal} {\bibinfo  {journal} {npj Quantum Inf}\ }\textbf {\bibinfo {volume}
  {7}},\ \bibinfo {pages} {129} (\bibinfo {year} {2021})}\BibitemShut {NoStop}%
\bibitem [{\citenamefont {Kim}\ \emph {et~al.}(2022)\citenamefont {Kim}, \citenamefont {Jünger}, \citenamefont {Morvan}, \citenamefont {Barnard}, \citenamefont {Livingston}, \citenamefont {Altoé}, \citenamefont {Kim}, \citenamefont {Song}, \citenamefont {Chen}, \citenamefont {Kreikebaum}, \citenamefont {Ogletree}, \citenamefont {Santiago},\ and\ \citenamefont {Siddiqi}}]{Kim_laser_annealing}%
  \BibitemOpen
  \bibfield  {author} {\bibinfo {author} {\bibfnamefont {H.}~\bibnamefont {Kim}}, \bibinfo {author} {\bibfnamefont {C.}~\bibnamefont {Jünger}}, \bibinfo {author} {\bibfnamefont {A.}~\bibnamefont {Morvan}}, \bibinfo {author} {\bibfnamefont {E.~S.}\ \bibnamefont {Barnard}}, \bibinfo {author} {\bibfnamefont {W.~P.}\ \bibnamefont {Livingston}}, \bibinfo {author} {\bibfnamefont {M.~V.~P.}\ \bibnamefont {Altoé}}, \bibinfo {author} {\bibfnamefont {Y.}~\bibnamefont {Kim}}, \bibinfo {author} {\bibfnamefont {C.}~\bibnamefont {Song}}, \bibinfo {author} {\bibfnamefont {L.}~\bibnamefont {Chen}}, \bibinfo {author} {\bibfnamefont {J.~M.}\ \bibnamefont {Kreikebaum}}, \bibinfo {author} {\bibfnamefont {D.~F.}\ \bibnamefont {Ogletree}}, \bibinfo {author} {\bibfnamefont {D.~I.}\ \bibnamefont {Santiago}},\ and\ \bibinfo {author} {\bibfnamefont {I.}~\bibnamefont {Siddiqi}},\ }\bibfield  {title} {\bibinfo {title} {Effects of laser-annealing on fixed-frequency superconducting qubits},\ }\href {https://doi.org/10.1063/5.0102092}
  {\bibfield  {journal} {\bibinfo  {journal} {Appl. Phys. Lett.}\ }\textbf {\bibinfo {volume} {121}},\ \bibinfo {pages} {142601} (\bibinfo {year} {2022})}\BibitemShut {NoStop}%
\bibitem [{\citenamefont {Sharma}\ \emph {et~al.}(2021)\citenamefont {Sharma}, \citenamefont {Elliott}, \citenamefont {Blomberg}, \citenamefont {Haukka}, \citenamefont {Givens}, \citenamefont {Tuominen},\ and\ \citenamefont {Ritala}}]{Sharma_Al2O3_fluorides}%
  \BibitemOpen
  \bibfield  {author} {\bibinfo {author} {\bibfnamefont {V.}~\bibnamefont {Sharma}}, \bibinfo {author} {\bibfnamefont {S.~D.}\ \bibnamefont {Elliott}}, \bibinfo {author} {\bibfnamefont {T.}~\bibnamefont {Blomberg}}, \bibinfo {author} {\bibfnamefont {S.}~\bibnamefont {Haukka}}, \bibinfo {author} {\bibfnamefont {M.~E.}\ \bibnamefont {Givens}}, \bibinfo {author} {\bibfnamefont {M.}~\bibnamefont {Tuominen}},\ and\ \bibinfo {author} {\bibfnamefont {M.}~\bibnamefont {Ritala}},\ }\bibfield  {title} {\bibinfo {title} {Thermal atomic layer etching of aluminum oxide ({$\text{Al}_2\text{O}_3$}) using sequential exposures of niobium pentafluoride ({$\text{NbF}_5$}) and carbon tetrachloride ({$\text{CCl}_4$}): A combined experimental and density functional theory study of the etch mechanism},\ }\href {https://doi.org/10.1021/acs.chemmater.1c00142} {\bibfield  {journal} {\bibinfo  {journal} {Chem. Mater.}\ }\textbf {\bibinfo {volume} {33}},\ \bibinfo {pages} {2883–2893} (\bibinfo {year} {2021})}\BibitemShut {NoStop}%
\bibitem [{\citenamefont {Roodenko}\ \emph {et~al.}(2010)\citenamefont {Roodenko}, \citenamefont {Seitz}, \citenamefont {Gogte}, \citenamefont {Veyan}, \citenamefont {Yan},\ and\ \citenamefont {Chabal}}]{Roodenko_XeF2_Al_effects}%
  \BibitemOpen
  \bibfield  {author} {\bibinfo {author} {\bibfnamefont {K.}~\bibnamefont {Roodenko}}, \bibinfo {author} {\bibfnamefont {O.}~\bibnamefont {Seitz}}, \bibinfo {author} {\bibfnamefont {Y.}~\bibnamefont {Gogte}}, \bibinfo {author} {\bibfnamefont {J.-F.}\ \bibnamefont {Veyan}}, \bibinfo {author} {\bibfnamefont {X.-M.}\ \bibnamefont {Yan}},\ and\ \bibinfo {author} {\bibfnamefont {Y.~J.}\ \bibnamefont {Chabal}},\ }\bibfield  {title} {\bibinfo {title} {Modification of the adhesive properties of {$\text{XeF}_2$}-etched aluminum surfaces by deposition of organic self-assembled monolayers},\ }\href {https://doi.org/10.1021/jp1068076} {\bibfield  {journal} {\bibinfo  {journal} {J. Phys. Chem. C}\ }\textbf {\bibinfo {volume} {114}},\ \bibinfo {pages} {22566–22572} (\bibinfo {year} {2010})}\BibitemShut {NoStop}%
\bibitem [{\citenamefont {Vepsäläinen}\ \emph {et~al.}(2020)\citenamefont {Vepsäläinen}, \citenamefont {Karamlou}, \citenamefont {Orrell}, \citenamefont {Dogra}, \citenamefont {Loer}, \citenamefont {Vasconcelos}, \citenamefont {Kim}, \citenamefont {Melville}, \citenamefont {Niedzielski}, \citenamefont {Yoder}, \citenamefont {Gustavsson}, \citenamefont {Formaggio}, \citenamefont {VanDevender},\ and\ \citenamefont {Oliver}}]{Vepsalainen_qp_ionizing_rad}%
  \BibitemOpen
  \bibfield  {author} {\bibinfo {author} {\bibfnamefont {A.~P.}\ \bibnamefont {Vepsäläinen}}, \bibinfo {author} {\bibfnamefont {A.~H.}\ \bibnamefont {Karamlou}}, \bibinfo {author} {\bibfnamefont {J.~L.}\ \bibnamefont {Orrell}}, \bibinfo {author} {\bibfnamefont {A.~S.}\ \bibnamefont {Dogra}}, \bibinfo {author} {\bibfnamefont {B.}~\bibnamefont {Loer}}, \bibinfo {author} {\bibfnamefont {F.}~\bibnamefont {Vasconcelos}}, \bibinfo {author} {\bibfnamefont {D.~K.}\ \bibnamefont {Kim}}, \bibinfo {author} {\bibfnamefont {A.~J.}\ \bibnamefont {Melville}}, \bibinfo {author} {\bibfnamefont {B.~M.}\ \bibnamefont {Niedzielski}}, \bibinfo {author} {\bibfnamefont {J.~L.}\ \bibnamefont {Yoder}}, \bibinfo {author} {\bibfnamefont {S.}~\bibnamefont {Gustavsson}}, \bibinfo {author} {\bibfnamefont {J.~A.}\ \bibnamefont {Formaggio}}, \bibinfo {author} {\bibfnamefont {B.~A.}\ \bibnamefont {VanDevender}},\ and\ \bibinfo {author} {\bibfnamefont {W.~D.}\ \bibnamefont {Oliver}},\ }\bibfield  {title} {\bibinfo {title} {Impact of
  ionizing radiation on superconducting qubit coherence},\ }\href {https://doi.org/10.1038/s41586-020-2619-8} {\bibfield  {journal} {\bibinfo  {journal} {Nature}\ }\textbf {\bibinfo {volume} {584}},\ \bibinfo {pages} {551–556} (\bibinfo {year} {2020})}\BibitemShut {NoStop}%
\bibitem [{\citenamefont {Jafri}\ \emph {et~al.}(1999)\citenamefont {Jafri}, \citenamefont {Busta},\ and\ \citenamefont {Walsh}}]{Jafri_CPD_1999}%
  \BibitemOpen
  \bibfield  {author} {\bibinfo {author} {\bibfnamefont {I.~H.}\ \bibnamefont {Jafri}}, \bibinfo {author} {\bibfnamefont {H.}~\bibnamefont {Busta}},\ and\ \bibinfo {author} {\bibfnamefont {S.~T.}\ \bibnamefont {Walsh}},\ }\bibfield  {title} {\bibinfo {title} {Critical point drying and cleaning for {MEMS} technology},\ }\bibfield  {journal} {\bibinfo  {journal} {Proc. SPIE 3880, MEMS Reliability for Critical and Space Applications}\ }\href {https://doi.org/10.1117/12.359371} {10.1117/12.359371} (\bibinfo {year} {1999})\BibitemShut {NoStop}%
\bibitem [{\citenamefont {Chistolini}\ \emph {et~al.}(2024)\citenamefont {Chistolini}, \citenamefont {Lee}, \citenamefont {Banerjee}, \citenamefont {Alghadeer}, \citenamefont {Jünger}, \citenamefont {Altoé}, \citenamefont {Song}, \citenamefont {Chen}, \citenamefont {Wang}, \citenamefont {Santiago},\ and\ \citenamefont {Siddiqi}}]{Chistolini_SiN_membrane}%
  \BibitemOpen
  \bibfield  {author} {\bibinfo {author} {\bibfnamefont {T.}~\bibnamefont {Chistolini}}, \bibinfo {author} {\bibfnamefont {K.}~\bibnamefont {Lee}}, \bibinfo {author} {\bibfnamefont {A.}~\bibnamefont {Banerjee}}, \bibinfo {author} {\bibfnamefont {M.}~\bibnamefont {Alghadeer}}, \bibinfo {author} {\bibfnamefont {C.}~\bibnamefont {Jünger}}, \bibinfo {author} {\bibfnamefont {M.~V.~P.}\ \bibnamefont {Altoé}}, \bibinfo {author} {\bibfnamefont {C.}~\bibnamefont {Song}}, \bibinfo {author} {\bibfnamefont {S.}~\bibnamefont {Chen}}, \bibinfo {author} {\bibfnamefont {F.}~\bibnamefont {Wang}}, \bibinfo {author} {\bibfnamefont {D.~I.}\ \bibnamefont {Santiago}},\ and\ \bibinfo {author} {\bibfnamefont {I.}~\bibnamefont {Siddiqi}},\ }\bibfield  {title} {\bibinfo {title} {Performance of superconducting resonators suspended on {SiN} membranes},\ }\href {https://doi.org/10.1063/5.0222680} {\bibfield  {journal} {\bibinfo  {journal} {Appl. Phys. Lett.}\ }\textbf {\bibinfo {volume} {125}},\ \bibinfo {pages} {204001} (\bibinfo
  {year} {2024})}\BibitemShut {NoStop}%
\bibitem [{\citenamefont {Koch}\ \emph {et~al.}(2007)\citenamefont {Koch}, \citenamefont {DiVincenzo},\ and\ \citenamefont {Clarke}}]{Koch_flux_noise}%
  \BibitemOpen
  \bibfield  {author} {\bibinfo {author} {\bibfnamefont {R.~H.}\ \bibnamefont {Koch}}, \bibinfo {author} {\bibfnamefont {D.~P.}\ \bibnamefont {DiVincenzo}},\ and\ \bibinfo {author} {\bibfnamefont {J.}~\bibnamefont {Clarke}},\ }\bibfield  {title} {\bibinfo {title} {Model for $1/f$ flux noise in squids and qubits},\ }\href {https://doi.org/10.1103/PhysRevLett.98.267003} {\bibfield  {journal} {\bibinfo  {journal} {Phys. Rev. Lett.}\ }\textbf {\bibinfo {volume} {98}},\ \bibinfo {pages} {267003} (\bibinfo {year} {2007})}\BibitemShut {NoStop}%
\bibitem [{\citenamefont {Kumar}\ \emph {et~al.}(2016)\citenamefont {Kumar}, \citenamefont {Sendelbach}, \citenamefont {Beck}, \citenamefont {Freeland}, \citenamefont {Wang}, \citenamefont {Wang}, \citenamefont {Yu}, \citenamefont {Wu}, \citenamefont {Pappas},\ and\ \citenamefont {McDermott}}]{Kumar_flux_noise}%
  \BibitemOpen
  \bibfield  {author} {\bibinfo {author} {\bibfnamefont {P.}~\bibnamefont {Kumar}}, \bibinfo {author} {\bibfnamefont {S.}~\bibnamefont {Sendelbach}}, \bibinfo {author} {\bibfnamefont {M.~A.}\ \bibnamefont {Beck}}, \bibinfo {author} {\bibfnamefont {J.~W.}\ \bibnamefont {Freeland}}, \bibinfo {author} {\bibfnamefont {Z.}~\bibnamefont {Wang}}, \bibinfo {author} {\bibfnamefont {H.}~\bibnamefont {Wang}}, \bibinfo {author} {\bibfnamefont {C.~C.}\ \bibnamefont {Yu}}, \bibinfo {author} {\bibfnamefont {R.~Q.}\ \bibnamefont {Wu}}, \bibinfo {author} {\bibfnamefont {D.~P.}\ \bibnamefont {Pappas}},\ and\ \bibinfo {author} {\bibfnamefont {R.}~\bibnamefont {McDermott}},\ }\bibfield  {title} {\bibinfo {title} {Origin and reduction of $1/f$ magnetic flux noise in superconducting devices},\ }\href {https://doi.org/10.1103/PhysRevApplied.6.041001} {\bibfield  {journal} {\bibinfo  {journal} {Phys. Rev. Appl.}\ }\textbf {\bibinfo {volume} {6}},\ \bibinfo {pages} {041001} (\bibinfo {year} {2016})}\BibitemShut {NoStop}%
\bibitem [{\citenamefont {Braum\"uller}\ \emph {et~al.}(2020)\citenamefont {Braum\"uller}, \citenamefont {Ding}, \citenamefont {Veps\"al\"ainen}, \citenamefont {Sung}, \citenamefont {Kjaergaard}, \citenamefont {Menke}, \citenamefont {Winik}, \citenamefont {Kim}, \citenamefont {Niedzielski}, \citenamefont {Melville}, \citenamefont {Yoder}, \citenamefont {Hirjibehedin}, \citenamefont {Orlando}, \citenamefont {Gustavsson},\ and\ \citenamefont {Oliver}}]{braumuller2020characterizing}%
  \BibitemOpen
  \bibfield  {author} {\bibinfo {author} {\bibfnamefont {J.}~\bibnamefont {Braum\"uller}}, \bibinfo {author} {\bibfnamefont {L.}~\bibnamefont {Ding}}, \bibinfo {author} {\bibfnamefont {A.~P.}\ \bibnamefont {Veps\"al\"ainen}}, \bibinfo {author} {\bibfnamefont {Y.}~\bibnamefont {Sung}}, \bibinfo {author} {\bibfnamefont {M.}~\bibnamefont {Kjaergaard}}, \bibinfo {author} {\bibfnamefont {T.}~\bibnamefont {Menke}}, \bibinfo {author} {\bibfnamefont {R.}~\bibnamefont {Winik}}, \bibinfo {author} {\bibfnamefont {D.}~\bibnamefont {Kim}}, \bibinfo {author} {\bibfnamefont {B.~M.}\ \bibnamefont {Niedzielski}}, \bibinfo {author} {\bibfnamefont {A.}~\bibnamefont {Melville}}, \bibinfo {author} {\bibfnamefont {J.~L.}\ \bibnamefont {Yoder}}, \bibinfo {author} {\bibfnamefont {C.~F.}\ \bibnamefont {Hirjibehedin}}, \bibinfo {author} {\bibfnamefont {T.~P.}\ \bibnamefont {Orlando}}, \bibinfo {author} {\bibfnamefont {S.}~\bibnamefont {Gustavsson}},\ and\ \bibinfo {author} {\bibfnamefont {W.~D.}\ \bibnamefont {Oliver}},\ }\bibfield
  {title} {\bibinfo {title} {Characterizing and optimizing qubit coherence based on squid geometry},\ }\href {https://doi.org/10.1103/PhysRevApplied.13.054079} {\bibfield  {journal} {\bibinfo  {journal} {Phys. Rev. Appl.}\ }\textbf {\bibinfo {volume} {13}},\ \bibinfo {pages} {054079} (\bibinfo {year} {2020})}\BibitemShut {NoStop}%
\bibitem [{\citenamefont {Rower}\ \emph {et~al.}(2023)\citenamefont {Rower}, \citenamefont {Ateshian}, \citenamefont {Li}, \citenamefont {Hays}, \citenamefont {Bluvstein}, \citenamefont {Ding}, \citenamefont {Kannan}, \citenamefont {Almanakly}, \citenamefont {Braum\"uller}, \citenamefont {Kim}, \citenamefont {Melville}, \citenamefont {Niedzielski}, \citenamefont {Schwartz}, \citenamefont {Yoder}, \citenamefont {Orlando}, \citenamefont {Wang}, \citenamefont {Gustavsson}, \citenamefont {Grover}, \citenamefont {Serniak}, \citenamefont {Comin},\ and\ \citenamefont {Oliver}}]{rower2023evolution}%
  \BibitemOpen
  \bibfield  {author} {\bibinfo {author} {\bibfnamefont {D.~A.}\ \bibnamefont {Rower}}, \bibinfo {author} {\bibfnamefont {L.}~\bibnamefont {Ateshian}}, \bibinfo {author} {\bibfnamefont {L.~H.}\ \bibnamefont {Li}}, \bibinfo {author} {\bibfnamefont {M.}~\bibnamefont {Hays}}, \bibinfo {author} {\bibfnamefont {D.}~\bibnamefont {Bluvstein}}, \bibinfo {author} {\bibfnamefont {L.}~\bibnamefont {Ding}}, \bibinfo {author} {\bibfnamefont {B.}~\bibnamefont {Kannan}}, \bibinfo {author} {\bibfnamefont {A.}~\bibnamefont {Almanakly}}, \bibinfo {author} {\bibfnamefont {J.}~\bibnamefont {Braum\"uller}}, \bibinfo {author} {\bibfnamefont {D.~K.}\ \bibnamefont {Kim}}, \bibinfo {author} {\bibfnamefont {A.}~\bibnamefont {Melville}}, \bibinfo {author} {\bibfnamefont {B.~M.}\ \bibnamefont {Niedzielski}}, \bibinfo {author} {\bibfnamefont {M.~E.}\ \bibnamefont {Schwartz}}, \bibinfo {author} {\bibfnamefont {J.~L.}\ \bibnamefont {Yoder}}, \bibinfo {author} {\bibfnamefont {T.~P.}\ \bibnamefont {Orlando}}, \bibinfo {author} {\bibfnamefont
  {J.~I.-J.}\ \bibnamefont {Wang}}, \bibinfo {author} {\bibfnamefont {S.}~\bibnamefont {Gustavsson}}, \bibinfo {author} {\bibfnamefont {J.~A.}\ \bibnamefont {Grover}}, \bibinfo {author} {\bibfnamefont {K.}~\bibnamefont {Serniak}}, \bibinfo {author} {\bibfnamefont {R.}~\bibnamefont {Comin}},\ and\ \bibinfo {author} {\bibfnamefont {W.~D.}\ \bibnamefont {Oliver}},\ }\bibfield  {title} {\bibinfo {title} {Evolution of $1/f$ flux noise in superconducting qubits with weak magnetic fields},\ }\href {https://doi.org/10.1103/PhysRevLett.130.220602} {\bibfield  {journal} {\bibinfo  {journal} {Phys. Rev. Lett.}\ }\textbf {\bibinfo {volume} {130}},\ \bibinfo {pages} {220602} (\bibinfo {year} {2023})}\BibitemShut {NoStop}%
\bibitem [{\citenamefont {Ambegaokar}\ and\ \citenamefont {Baratoff}(1963)}]{Ambegaokar_tunneling}%
  \BibitemOpen
  \bibfield  {author} {\bibinfo {author} {\bibfnamefont {V.}~\bibnamefont {Ambegaokar}}\ and\ \bibinfo {author} {\bibfnamefont {A.}~\bibnamefont {Baratoff}},\ }\bibfield  {title} {\bibinfo {title} {Tunneling between superconductors},\ }\href {https://doi.org/10.1103/PhysRevLett.10.486} {\bibfield  {journal} {\bibinfo  {journal} {Phys. Rev. Lett.}\ }\textbf {\bibinfo {volume} {10}},\ \bibinfo {pages} {486} (\bibinfo {year} {1963})}\BibitemShut {NoStop}%
\bibitem [{\citenamefont {Weides}\ \emph {et~al.}(2011)\citenamefont {Weides}, \citenamefont {Bialczak}, \citenamefont {Lenander}, \citenamefont {Lucero}, \citenamefont {Mariantoni}, \citenamefont {Neeley}, \citenamefont {O'Connell}, \citenamefont {Sank}, \citenamefont {Wang},\ and\ \citenamefont {Wenner}}]{Weides_phase_qb_fab}%
  \BibitemOpen
  \bibfield  {author} {\bibinfo {author} {\bibfnamefont {M.}~\bibnamefont {Weides}}, \bibinfo {author} {\bibfnamefont {R.~C.}\ \bibnamefont {Bialczak}}, \bibinfo {author} {\bibfnamefont {M.}~\bibnamefont {Lenander}}, \bibinfo {author} {\bibfnamefont {E.}~\bibnamefont {Lucero}}, \bibinfo {author} {\bibfnamefont {M.}~\bibnamefont {Mariantoni}}, \bibinfo {author} {\bibfnamefont {M.}~\bibnamefont {Neeley}}, \bibinfo {author} {\bibfnamefont {A.~D.}\ \bibnamefont {O'Connell}}, \bibinfo {author} {\bibfnamefont {D.}~\bibnamefont {Sank}}, \bibinfo {author} {\bibfnamefont {H.}~\bibnamefont {Wang}},\ and\ \bibinfo {author} {\bibfnamefont {J.}~\bibnamefont {Wenner}},\ }\bibfield  {title} {\bibinfo {title} {Phase qubits fabricated with trilayer junctions},\ }\href {https://doi.org/10.1088/0953-2048/24/5/055005} {\bibfield  {journal} {\bibinfo  {journal} {Supercond. Sci. Technol.}\ }\textbf {\bibinfo {volume} {24}},\ \bibinfo {pages} {055005} (\bibinfo {year} {2011})}\BibitemShut {NoStop}%
\bibitem [{\citenamefont {Zhu}\ \emph {et~al.}(2013)\citenamefont {Zhu}, \citenamefont {Ferguson}, \citenamefont {Manucharyan},\ and\ \citenamefont {Koch}}]{zhu2013circuit}%
  \BibitemOpen
  \bibfield  {author} {\bibinfo {author} {\bibfnamefont {G.}~\bibnamefont {Zhu}}, \bibinfo {author} {\bibfnamefont {D.~G.}\ \bibnamefont {Ferguson}}, \bibinfo {author} {\bibfnamefont {V.~E.}\ \bibnamefont {Manucharyan}},\ and\ \bibinfo {author} {\bibfnamefont {J.}~\bibnamefont {Koch}},\ }\bibfield  {title} {\bibinfo {title} {Circuit qed with fluxonium qubits: Theory of the dispersive regime},\ }\href {https://doi.org/10.1103/PhysRevB.87.024510} {\bibfield  {journal} {\bibinfo  {journal} {Phys. Rev. B}\ }\textbf {\bibinfo {volume} {87}},\ \bibinfo {pages} {024510} (\bibinfo {year} {2013})}\BibitemShut {NoStop}%
\bibitem [{\citenamefont {Nguyen}\ \emph {et~al.}(2022)\citenamefont {Nguyen}, \citenamefont {Koolstra}, \citenamefont {Kim}, \citenamefont {Morvan}, \citenamefont {Chistolini}, \citenamefont {Singh}, \citenamefont {Nesterov}, \citenamefont {Jünger}, \citenamefont {Chen}, \citenamefont {Pedramrazi}, \citenamefont {Mitchell}, \citenamefont {Kreikebaum}, \citenamefont {Puri}, \citenamefont {Santiago},\ and\ \citenamefont {Siddiqi}}]{Nguyen_blueprint_fluxonium}%
  \BibitemOpen
  \bibfield  {author} {\bibinfo {author} {\bibfnamefont {L.~B.}\ \bibnamefont {Nguyen}}, \bibinfo {author} {\bibfnamefont {G.}~\bibnamefont {Koolstra}}, \bibinfo {author} {\bibfnamefont {Y.}~\bibnamefont {Kim}}, \bibinfo {author} {\bibfnamefont {A.}~\bibnamefont {Morvan}}, \bibinfo {author} {\bibfnamefont {T.}~\bibnamefont {Chistolini}}, \bibinfo {author} {\bibfnamefont {S.}~\bibnamefont {Singh}}, \bibinfo {author} {\bibfnamefont {K.~N.}\ \bibnamefont {Nesterov}}, \bibinfo {author} {\bibfnamefont {C.}~\bibnamefont {Jünger}}, \bibinfo {author} {\bibfnamefont {L.}~\bibnamefont {Chen}}, \bibinfo {author} {\bibfnamefont {Z.}~\bibnamefont {Pedramrazi}}, \bibinfo {author} {\bibfnamefont {B.~K.}\ \bibnamefont {Mitchell}}, \bibinfo {author} {\bibfnamefont {J.~M.}\ \bibnamefont {Kreikebaum}}, \bibinfo {author} {\bibfnamefont {S.}~\bibnamefont {Puri}}, \bibinfo {author} {\bibfnamefont {D.~I.}\ \bibnamefont {Santiago}},\ and\ \bibinfo {author} {\bibfnamefont {I.}~\bibnamefont {Siddiqi}},\ }\bibfield  {title} {\bibinfo
  {title} {Blueprint for a high-performance fluxonium quantum processor},\ }\href {https://doi.org/10.1103/PRXQuantum.3.037001} {\bibfield  {journal} {\bibinfo  {journal} {PRX Quantum}\ }\textbf {\bibinfo {volume} {3}},\ \bibinfo {pages} {037001} (\bibinfo {year} {2022})}\BibitemShut {NoStop}%
\bibitem [{\citenamefont {Rigetti}\ \emph {et~al.}(2012)\citenamefont {Rigetti}, \citenamefont {Gambetta}, \citenamefont {Poletto}, \citenamefont {Plourde}, \citenamefont {Chow}, \citenamefont {C\'orcoles}, \citenamefont {Smolin}, \citenamefont {Merkel}, \citenamefont {Rozen}, \citenamefont {Keefe}, \citenamefont {Rothwell}, \citenamefont {Ketchen},\ and\ \citenamefont {Steffen}}]{rigetti2012superconducting}%
  \BibitemOpen
  \bibfield  {author} {\bibinfo {author} {\bibfnamefont {C.}~\bibnamefont {Rigetti}}, \bibinfo {author} {\bibfnamefont {J.~M.}\ \bibnamefont {Gambetta}}, \bibinfo {author} {\bibfnamefont {S.}~\bibnamefont {Poletto}}, \bibinfo {author} {\bibfnamefont {B.~L.~T.}\ \bibnamefont {Plourde}}, \bibinfo {author} {\bibfnamefont {J.~M.}\ \bibnamefont {Chow}}, \bibinfo {author} {\bibfnamefont {A.~D.}\ \bibnamefont {C\'orcoles}}, \bibinfo {author} {\bibfnamefont {J.~A.}\ \bibnamefont {Smolin}}, \bibinfo {author} {\bibfnamefont {S.~T.}\ \bibnamefont {Merkel}}, \bibinfo {author} {\bibfnamefont {J.~R.}\ \bibnamefont {Rozen}}, \bibinfo {author} {\bibfnamefont {G.~A.}\ \bibnamefont {Keefe}}, \bibinfo {author} {\bibfnamefont {M.~B.}\ \bibnamefont {Rothwell}}, \bibinfo {author} {\bibfnamefont {M.~B.}\ \bibnamefont {Ketchen}},\ and\ \bibinfo {author} {\bibfnamefont {M.}~\bibnamefont {Steffen}},\ }\bibfield  {title} {\bibinfo {title} {Superconducting qubit in a waveguide cavity with a coherence time approaching 0.1 ms},\ }\href
  {https://doi.org/10.1103/PhysRevB.86.100506} {\bibfield  {journal} {\bibinfo  {journal} {Phys. Rev. B}\ }\textbf {\bibinfo {volume} {86}},\ \bibinfo {pages} {100506} (\bibinfo {year} {2012})}\BibitemShut {NoStop}%
\bibitem [{\citenamefont {Yan}\ \emph {et~al.}(2018)\citenamefont {Yan}, \citenamefont {Campbell}, \citenamefont {Krantz}, \citenamefont {Kjaergaard}, \citenamefont {Kim}, \citenamefont {Yoder}, \citenamefont {Hover}, \citenamefont {Sears}, \citenamefont {Kerman}, \citenamefont {Orlando}, \citenamefont {Gustavsson},\ and\ \citenamefont {Oliver}}]{yan2018distinguishing}%
  \BibitemOpen
  \bibfield  {author} {\bibinfo {author} {\bibfnamefont {F.}~\bibnamefont {Yan}}, \bibinfo {author} {\bibfnamefont {D.}~\bibnamefont {Campbell}}, \bibinfo {author} {\bibfnamefont {P.}~\bibnamefont {Krantz}}, \bibinfo {author} {\bibfnamefont {M.}~\bibnamefont {Kjaergaard}}, \bibinfo {author} {\bibfnamefont {D.}~\bibnamefont {Kim}}, \bibinfo {author} {\bibfnamefont {J.~L.}\ \bibnamefont {Yoder}}, \bibinfo {author} {\bibfnamefont {D.}~\bibnamefont {Hover}}, \bibinfo {author} {\bibfnamefont {A.}~\bibnamefont {Sears}}, \bibinfo {author} {\bibfnamefont {A.~J.}\ \bibnamefont {Kerman}}, \bibinfo {author} {\bibfnamefont {T.~P.}\ \bibnamefont {Orlando}}, \bibinfo {author} {\bibfnamefont {S.}~\bibnamefont {Gustavsson}},\ and\ \bibinfo {author} {\bibfnamefont {W.~D.}\ \bibnamefont {Oliver}},\ }\bibfield  {title} {\bibinfo {title} {Distinguishing coherent and thermal photon noise in a circuit quantum electrodynamical system},\ }\href {https://doi.org/10.1103/PhysRevLett.120.260504} {\bibfield  {journal} {\bibinfo
  {journal} {Phys. Rev. Lett.}\ }\textbf {\bibinfo {volume} {120}},\ \bibinfo {pages} {260504} (\bibinfo {year} {2018})}\BibitemShut {NoStop}%
\end{thebibliography}
\end{document}